\documentclass[aps,prl,floatfix,twocolumn,showpacs,nofitinbib,hyperref=pdftex,citeautoscript,superscriptaddress]{revtex4-1}

\usepackage{epsfig}
\usepackage{amsmath}
\usepackage{amssymb}
\usepackage{amsfonts}
\usepackage{ dsfont }
\usepackage{ comment,color }
\usepackage{lipsum}
\usepackage{hyperref,mathtools}
\usepackage{makecell}
\usepackage{array}
\usepackage{placeins}
\usepackage{pifont}
\usepackage{siunitx}
\hypersetup{colorlinks=true,linkcolor=blue,anchorcolor=blue,citecolor=blue,filecolor=blue,urlcolor=blue,bookmarksnumbered=true,pdfview=FitB}

\newcommand{\beqa}{\begin{eqnarray}}
\newcommand{\eeqa}{\end{eqnarray}}

\begin{document}

\hsize\textwidth\columnwidth\hsize\csname@twocolumnfalse\endcsname

\title{Supercurrent in the presence of direct transmission and a resonant localized state}

\author{Vukan Levajac}
\address{QuTech and Kavli Institute of Nanoscience, Delft University of Technology, 2600 GA Delft, The Netherlands}

\author{Hristo Barakov}
\address{Kavli Institute of Nanoscience, Delft University of Technology, 2628 CJ Delft, The Netherlands}

\author{Grzegorz P. Mazur}
\author{Nick van Loo}
\author{Leo P. Kouwenhoven}
\address{QuTech and Kavli Institute of Nanoscience, Delft University of Technology, 2600 GA Delft, The Netherlands}

\author{Yuli V. Nazarov}
\email{y.v.nazarov@tudelft.nl}
\address{Kavli Institute of Nanoscience, Delft University of Technology, 2628 CJ Delft, The Netherlands}

\author{Ji-Yin Wang}
\email{wangjiyinshu@gmail.com}
\address{QuTech and Kavli Institute of Nanoscience, Delft University of Technology, 2600 GA Delft, The Netherlands}
\address{Beijing Academy of Quantum Information Sciences, 100193 Beijing, China}

\date{\today}

\vskip1.5truecm
\begin{abstract}
We study the current-phase relation (CPR) of an InSb-Al nanowire Josephson junction in parallel magnetic fields up to $700$\,mT. At high magnetic fields and in narrow voltage intervals of a gate under the junction, the CPR exhibits $\pi$-shifts. The supercurrent declines within these gate intervals and shows asymmetric gate voltage dependence above and below them. We detect these features sometimes also at zero magnetic field. The observed CPR properties are reproduced by a theoretical model of supercurrent transport via interference between direct transmission and a resonant localized state.
\end{abstract}


\maketitle

A Josephson junction (JJ) consists of two superconductors (S) and a weak link between them that supports transport of Cooper pairs in the form of a non-dissipative supercurrent \cite{pi_josephson_physlett_1962}. If the weak link is a normal conductor (N), Andreev reflections at the two SN interfaces give rise to Andreev levels inside the junction that mediate the supercurrent \cite{pi_prada_natrev_2020}. JJs with semiconducting weak links are widely used to study the influence of their tunable properties on the Andreev spectrum and supercurrent. This is evident in multiple superconducting phenomena, such as topological superconductivity \cite{pi_pientka_prx_2017, pi_hell_prl_2017, pi_schrade_prb_2017, pi_cayao_prb_2017, pi_schrade_prl_2018, pi_fornieri_nat_2019, pi_ren_nat_2019, pi_dartialih_prl_2021}, the anomalous Josephson effect \cite{pi_yokoyama_jphyssocjpn_2013, pi_yokoyama_prb_2014, pi_szombati_natphys_2016, pi_strambini_natnano_2020, pi_wang_sciadv_2022} and the Josephson diode effect \cite{pi_chen_prb_2018, pi_legg_prb_2022, pi_souto_prl_2022, pi_wu_nat_2022, pi_baumgartmer_natnano_, pi_turini_nanolett_2022, pi_vanloo_arxiv_diode_2021, pi_zhang_arxiv_2022}. Semiconducting JJs also have attractive applications in quantum computing, such as gate-tunable superconducting qubits \cite{pi_delange_prl-2015, pi_larsen_prl_2015, pi_casparis_natnano_2018} and Andreev spin qubits \cite{pi_zazunov_prl_2003, pi_chtchelkatchev_prl_2003, pi_janvier_science_2015, pi_hays_natphys_2020, pi_hays_science_2021, pi_pitavidal_natphys_2023}.

A $\pi$-shift in the current-phase relation (CPR) of a JJ can occur due to spin-splitting of the Andreev levels in an external magnetic field. Once the splitting is of the order of the superconducting gap, the minimum of the ground state energy moves from the $0$-phase to the $\pi$-phase, and the junction undergoes a $0-\pi$ transition \cite{pi_yokoyama_jphyssocjpn_2013, pi_yokoyama_prb_2014}. In the presence of Coulomb interaction, sequential coherent single-electron tunneling can result in a supercurrent which direction depends on the junction parity, and $0-\pi$ transitions can occur even at zero magnetic field \cite{pi_spivak_prb_1991,pi_vecino_pr_2003, pi_meng_prb_2009,pi_lee_pr_2022}. These phenomena have been studied for quantum dot-based JJs made from semiconducting nanowires (NWs) \cite{pi_vandam_nature_2006, pi_jorgensen_nanolett_2007, pi_lee_natnano_2014, pi_li_prb_2017, pi_bargerbos_prxq_2022}, carbon nanotubes (CNTs) \cite{pi_eichler_prb_2009, pi_maurand_prx_2012, pi_kim_prl_2013} and two-dimensional electron gases (2DEGs) \cite{pi_whiticar_prb_2021}. 

In the above works that study JJs in hybrid semiconductor-superconductor nanowires, the CPR has been measured only at low parallel magnetic fields (up to several tens of $\mathrm{mT}$). Studying CPR of nanowire JJs in high parallel fields is motivated by various proposals for detecting signatures of a topological phase transition in supercurrent measurements \cite{pi_schrade_prb_2017, pi_schrade_prl_2018, pi_sanjose_prl_2014, pi_cxliu_prb_2021}.  

In this Letter, we embed two hybrid InSb-Al nanowire JJs into a superconducting quantum interference device (SQUID) and we study the CPR of one JJ at parallel magnetic fields that are unprecedentedly high for Al-based nanowire JJs - exceeding $700\,\mathrm{mT}$. At zero magnetic field, localized states in the junction are identified by observing resonances in the normal-state conductance. When the localized states become involved in the superconducting transport as they are tuned close to the Fermi energy by a gate under the junction, the supercurrent exhibits asymmetric amplitude modulation by the gate. In these gate intervals at zero magnetic field, either pairs of $0-\pi$ transitions give rise to $\pi$-regions in the CPR, or the supercurrent is enhanced with constant phase. We further investigate the CPR of the nanowire JJ by increasing parallel magnetic field and find that high fields can enlarge $\pi$-regions or drive $0-\pi$ transitions in the CPR. In order to understand these phenomena, we develop a model involving a transmission channel and a localized resonant state inside a single nanowire JJ. This model can reproduce the main interesting features observed in the experiment by considering the interference between the transmission channel and the localized state. This interference effect of supercurrent represents a novel, superconducting version of the  well-known Fano effect \cite{pi_fano_pr_1961}.


\begin{figure}
\centering
\includegraphics[width=1\linewidth]{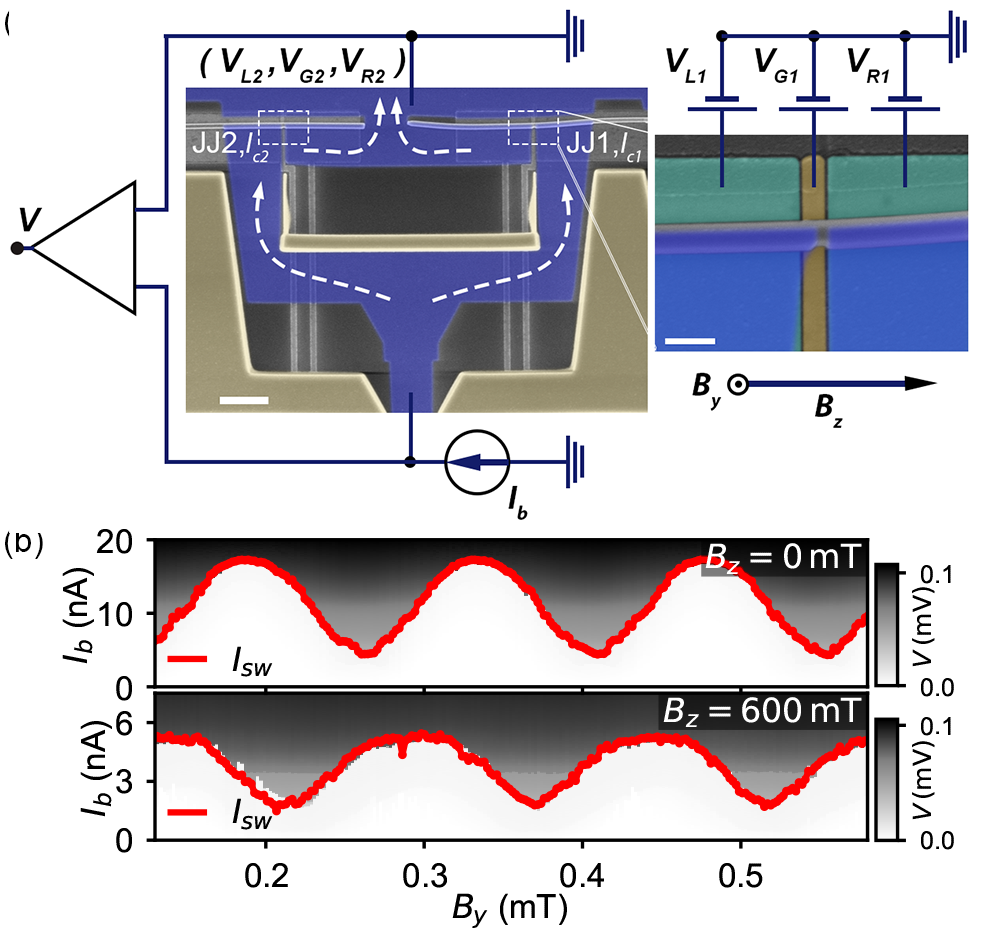}
\caption{\textbf{(a)} False-colored SEM image of the nanowire SQUID (left, scale bar $1\,\mathrm{\mu m}$) and a zoom-in on one arm (right, scale bar $100\,\mathrm{nm}$). Dielectric shadow-walls (yellow) define the two Josephson junctions (JJ1 and JJ2 with critical currents $I_{c1}$ and $I_{c2}$) in two nearly parallel InSb nanowires, and the superconducting Al (blue) loop on the substrate. Bias current $I_b$ is applied and voltage drop $V$ is measured between the source and drain of the SQUID. White arrows indicate the current directions along the two SQUID arms. JJ1 and JJ2 have $\sim 40\,\mathrm{nm}$-long junctions and are coupled via $20\,\mathrm{nm}$ of $\mathrm{HfO_2}$ to corresponding three underlying gates at voltages $V_{Li}$, $V_{Gi}$ and $V_{Ri}$ (i=1,2). Magnetic fields $B_z$ and $B_y$ are applied parallel to JJ1 and perpendicular to the SQUID loop, respectively. \textbf{(b)} $V$ as a function of $I_b$ and $B_y$ at $B_z=0\,\mathrm{mT}$ (top) and $B_z=600\,\mathrm{mT}$ (bottom). The red traces $I_{sw}$ are obtained in a setup for fast switching current measurements. The gate settings are $V_{G1}=2.36\,\mathrm{V}$, $V_{L1}=V_{R1}=1.5\,\mathrm{V}$, $V_{G2}=2.88\,\mathrm{V}$ and $V_{L2}=V_{R2}=1.75\,\mathrm{V}$.}\label{fig:pi_1}
\end{figure}

The nanowire SQUID is introduced in Fig. \ref{fig:pi_1}. In Fig. \ref{fig:pi_1}(a) on the left, a false-colored scanning electron microscopy (SEM) image displays two InSb-Al nanowire Josephson junctions - JJ1 and JJ2 - enclosed in a superconducting Al (blue) loop. The Al layout is obtained through the shadow-wall (yellow) lithography \cite{pi_heedt_natcomm_2021, pi_borsoi_afm_2021, pi_levajac_arxiv_2022}. White arrows indicate the current paths via the two SQUID arms between the source and drain that are shared by JJ1 and JJ2. A zoom-in on JJ1 is displayed in Fig. \ref{fig:pi_1}(a) on the right. The junction has a length of $\sim 40\,\mathrm{nm}$ and its electro-chemical potential is controlled by an underlying gate with a voltage $V_{G1}$. The other two underlying gates with voltages $V_{L1}$ and $V_{R1}$ predominantly tune the electro-chemical potential in the nanowire sections covered by the left and right lead. JJ2 has nominally the same design as JJ1. An in-plane magnetic field $B_z$ is applied parallel to the nanowire of JJ1 and the flux through the loop is introduced by an out-of-plane magnetic field $B_y$. For more details of the device fabrication, see our recent work \cite{pi_levajac_arxiv_2022} where the identical nanowire SQUID has been introduced.

We characterize the SQUID in the standard four-terminal setup where a bias current $I_b$ is applied and a voltage drop $V$ is measured between the source and drain. $V-I_b$ traces are measured while the flux through the loop is swept by varying $B_y$. The measurement is performed at $B_z=0\,\mathrm{mT}$ and $B_z=600\,\mathrm{mT}$ (see Fig. \ref{fig:pi_1}(b)). Aside from the measurements shown in the 2D-maps, a setup for switching current measurements in a fast way is employed \cite{pi_vanloo_arxiv_diode_2021, pi_vanwoerkom_natphys_2015}. In this setup, $I_b$ is ramped and $V$ is monitored without recording the $V-I_b$ trace. For each $I_b$-ramp, only a single switching current value is recorded as the $I_b$ value for which the SQUID switches from the superconducting to the resistive regime - as $V$ crosses a pre-defined threshold voltage ($7\,\mathrm{\mu V}$ in our measurements) \cite{sm}. Switching current $I_{sw}$ is obtained as an average of five switching current values measured consecutively upon setting $B_y$. Such obtained $I_{sw}(B_y)$ dependences (red traces in Fig. \ref{fig:pi_1}(b)) overlap well with the oscillatory boundaries between the superconducting and resistive regime - demonstrating the accuracy of the fast switching current measurements. The SQUID oscillations at high $B_z$ confirm the resilience of supercurrent interference against large magnetic fields \cite{pi_levajac_arxiv_2022}. In the rest of this work, we employ the fast measurement setup to obtain switching current of the SQUID. At zero magnetic field, we use the gates under JJ2 to tune its critical current so that $I_{c2} \gg I_{c1}$. In such a highly asymmetric SQUID configuration, JJ2 serves as the reference arm and the CPR of JJ1 is directly obtained by measuring an $I_{sw}(B_y)$ trace. In our SQUID, parallel magnetic fields simultaneously suppress supercurrent in both JJs, and the condition $I_{c2} \gg I_{c1}$ may not always be satisfied at high $B_z$. In spite of this, the CPR of JJ1 could still be reflected in the SQUID oscillations. For example, a 0-$\pi$ transition in JJ1 would cause a half-period shift in the $I_{sw}(B_y)$ trace. 

\begin{figure*} 
\includegraphics[width=1\linewidth] {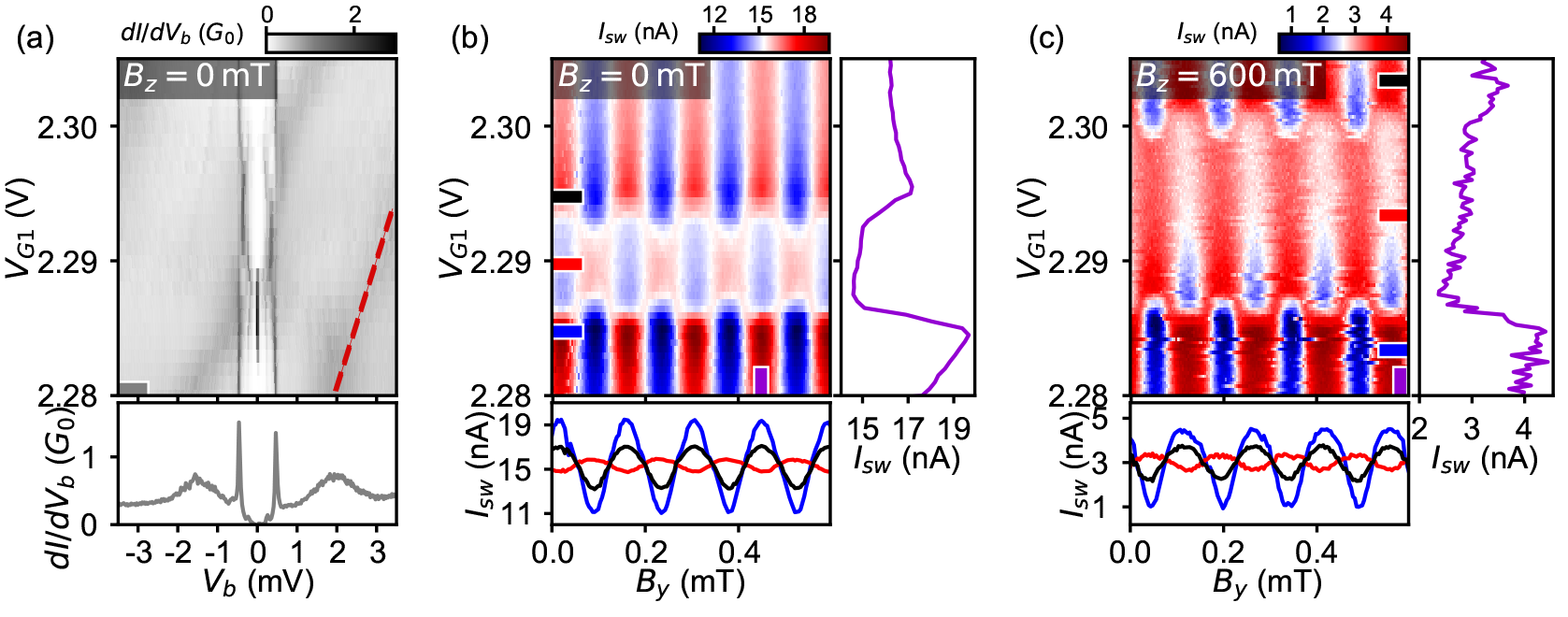}
\caption{\textbf{(a)} Differential conductance $dI/dV_b$ of JJ1 as a function of $V_b$ and $V_{G1}$ at zero magnetic field and $V_{L1}=V_{R1}=1.5\,\mathrm{V}$. JJ2 is pinched-off by setting $V_{G2}=V_{L2}=V_{R2}=0\,\mathrm{V}$. A linecut (grey marker) with sharp coherence peaks at $\pm2\Delta/e$ and broad resonance peaks is shown below. One resonance is marked by a red dashed line and the gate lever arm $\alpha \sim 0.5\Delta/\mathrm{mV}$ is estimated from its slope in the $(V_b,V_{G1})$-plane. \textbf{(b)} $I_{sw}$ as a function of $B_y$ and $V_{G1}$ at $B_z=0\,\mathrm{mT}$ and $V_{L1}=V_{R1}=1.5\,\mathrm{V}$. JJ2 is turned on by setting $V_{G2}=2.88\,\mathrm{V}$ and $V_{L2}=V_{R2}=1.75\,\mathrm{V}$. Horizontal linecuts (red, blue and black marker) are shown below and a vertical linecut (purple marker) is displayed on the right. \textbf{(c)} Same as (b), but for $B_z=600\,\mathrm{mT}$.}\label{fig:pi_2}
\end{figure*}

We first pinch-off the reference arm and 
measure the differential conductance $dI/dV_b$ of JJ1 in a two-terminal setup with the standard lock-in configuration \cite{sm}. In Fig. \ref{fig:pi_2}(a), $dI/dV_b$ at zero magnetic field is shown as a function of a bias voltage $V_b$ and $V_{G1}$. By finding the positions of the coherence peaks in the grey linecut, the superconducting gap of the leads $\Delta \sim 0.23\,\mathrm{meV}$ is extracted. Conductance peaks at $0<|V_b|<2\Delta/e$ and a zero-bias peak (ZBP) in the 2D-map correspond to the superconducting transport via multiple Andreev reflections (MARs) and supercurrent, respectively. Next to this, in the normal-state transport for $|V_b|>2\Delta/e$, resonances with positive slope in the $(V_b,V_{G1})$-plane are observed and one resonance is marked by a red dashed line. The resonances indicate the presence of a state that is localized in the junction and coupled to both leads. Coupling to the leads $\Gamma$ and charging energy $U$ attributed to the localized state are estimated from the two resonance peaks for $|V_b|>2\Delta/e$ (grey linecut). From the FWHM of the resonance peaks, the coupling is estimated as $\Gamma \sim 1\,\mathrm{meV} \sim 4\Delta$. Since no resonances with negative slope in the $(V_b,V_{G1})$-plane are visible, only an upper limit for $U$ can be estimated as the separation of two neighbouring resonance peaks along the $V_b$-axis - giving $U<4\,\mathrm{meV} \sim 16\Delta$. The features of superconducting transport exhibit strong modulation as the localized state approaches the in-gap energies and contributes to the superconducting transport. In order to investigate the influence of the localized state on the supercurrent, we turn on the reference arm and use the SQUID configuration to investigate the CPR. 

Fig. \ref{fig:pi_2}(b) displays $I_{sw}$ of the SQUID as a function of $B_y$ and $V_{G1}$ in the voltage range studied in Fig. \ref{fig:pi_2}(a). Three distinct regions can be identified - with the middle region being $\pi$-shifted ($\pi$-region) relative to the regions below and above ($0$-regions). Noticeably, the $\pi$-region occurs in the same $V_{G1}$ interval in which the localized state is tuned below the superconducting gap (see Fig. \ref{fig:pi_2}(a)). Two horizontal linecuts in the two $0$-regions (blue and black) and one horizontal linecut in the middle of the $\pi$-region (red) demonstrates that the supercurrent declines inside the $\pi$-region - as the red linecut has the smallest amplitude. The blue and black linecut - taken symmetrically with respect to the red linecut - yet have very different amplitudes. This indicates an asymmetric gate voltage dependence of the supercurrent below and above the $\pi$-region. This asymmetry is further confirmed by a vertical linecut (purple). Upon increasing the parallel field to $B_z=600\,\mathrm{mT}$, $I_{sw}$ is measured in the same $B_y$ and $V_{G1}$ ranges and the result is shown in Fig. \ref{fig:pi_2}(c). Four linecuts taken analogously as in Fig. \ref{fig:pi_2}(b) demonstrate that the supercurrent suppression inside the $\pi$-region and the asymmetry between the $0$-regions remain at $B_z=600\,\mathrm{mT}$. However, both effects are less prominent in the high parallel field. In addition, the high magnetic field causes a broadening of the $\pi$-region along the gate voltage axis \cite{pi_bargerbos_prxq_2022}. The observed expansion of $\sim 7.5\,\mathrm{mV}$ in $V_{G1}$ corresponds to an energy of $\sim 0.9\,\mathrm{meV} \sim 4\Delta$ (see Fig. \ref{fig:pi_2}(a) for the estimation of the gate lever arm) and a same Zeeman energy $g \mu_B B \sim 0.9\,\mathrm{meV}$ would yield a $g$-factor $g \sim 26$. 

\begin{figure}
\centering
\includegraphics[width=1\linewidth]{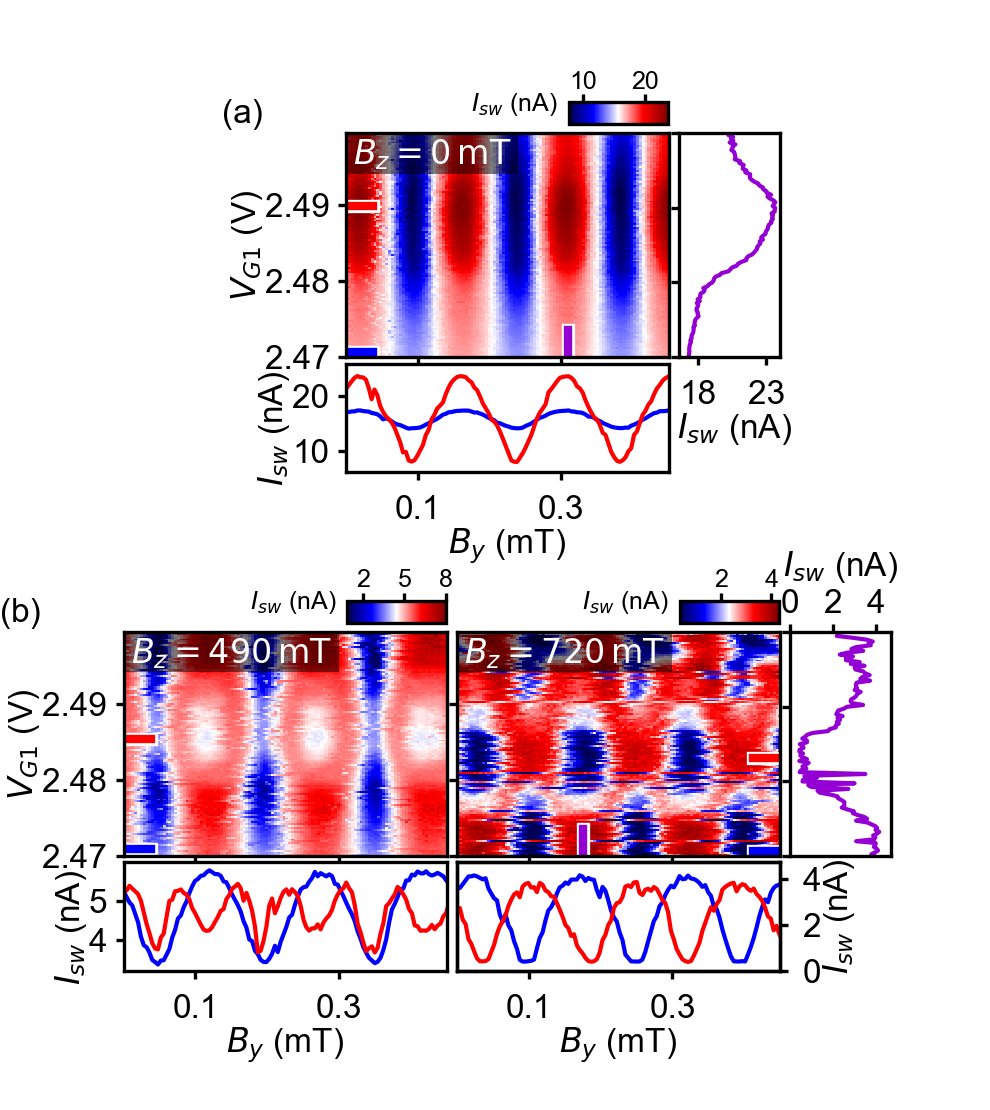}
\caption{$I_{sw}$ as a function of $B_y$ and $V_{G1}$ at: \textbf{(a)} $B_z=0\,\mathrm{mT}$ and \textbf{(b)} $B_z=490\,\mathrm{mT}$ (left) and $B_z=720\,\mathrm{mT}$ (right). The other gate voltages are $V_{L1}=V_{R1}=1.75\,\mathrm{V}$, $V_{G2}=2.88\,\mathrm{V}$ and $V_{L2}=V_{R2}=1.75\,\mathrm{V}$. Two linecuts (red and blue markers) are shown below and a vertical linecut (purple marker) is shown for $B_z=0\,\mathrm{mT}$ and $B_z=720\,\mathrm{mT}$ on the right. Three sharp peaks in the purple linecut at high $B_z$ are caused by instabilities of the flux.
}\label{fig:pi_3}
\end{figure}

We proceed by studying the CPR in another gate voltage interval, that is above the one presented in Fig. \ref{fig:pi_2}. In Fig. \ref{fig:pi_3}(a), $I_{sw}$ dependence on $B_y$ and $V_{G1}$ at $B_z=0\,\mathrm{mT}$ is presented, with two horizontal linecuts (red and blue) and one vertical linecut (purple). The linecuts show that the supercurrent is enhanced by the gate around $V_{G1}=2.49\,\mathrm{V}$. Moreover, the purple linecut demonstrates that the supercurrent amplitude is asymmetrically modulated below and above the enhancement. In contrast to Fig. \ref{fig:pi_2}(b), the asymmetric modulation in this $V_{G1}$ range is not accompanied by $0-\pi$ transitions - as evident from the horizontal linecuts. Next, we increase $B_z$ and measure $I_{sw}$ in the same $B_y$ and $V_{G1}$ ranges - at $B_z=490\,\mathrm{mT}$ and $B_z=720\,\mathrm{mT}$ (see Fig. \ref{fig:pi_3}(b)). At $B_z=720\,\mathrm{mT}$, a $\pi$-region is observed in the studied $V_{G1}$ interval and the supercurrent amplitudes inside the $\pi$-region (red linecut) and $0$-region (blue linecut) are comparable, which indicates a very weak suppression of the supercurrent inside the $\pi$-region. In addition, the asymmetry between the two $0$-regions is also very weak (purple linecut). We see that the $I_{sw}(B_y)$ traces are not sinusoidal any more, with $I_{sw}$ approaching zero at specific $B_y$ values. This indicates that the supercurrent amplitudes in the two SQUID arms are comparable. In spite of this, the $\pi$-region could still be well identified in the figure. At $B_z=490\,\mathrm{mT}$, the frequency of $I_{sw}$ oscillations doubles (red linecut) in a narrow $V_{G1}$ interval. Such a double frequency oscillation has been studied in theory \cite{pi_yokoyama_prb_2014,pi_vecino_pr_2003,pi_lee_pr_2022} and observed in experiment \cite{pi_bargerbos_prxq_2022}, and is due to an intermediate regime existing in-between a stable 0- and $\pi$-region. Therefore, the $I_{sw}$ dependence at $B_z=490\,\mathrm{mT}$ could be understood as an intermediate regime before a stable $\pi$ phase is formed at higher magnetic fields. 

\begin{figure}
\centering
\includegraphics[width=1\linewidth]{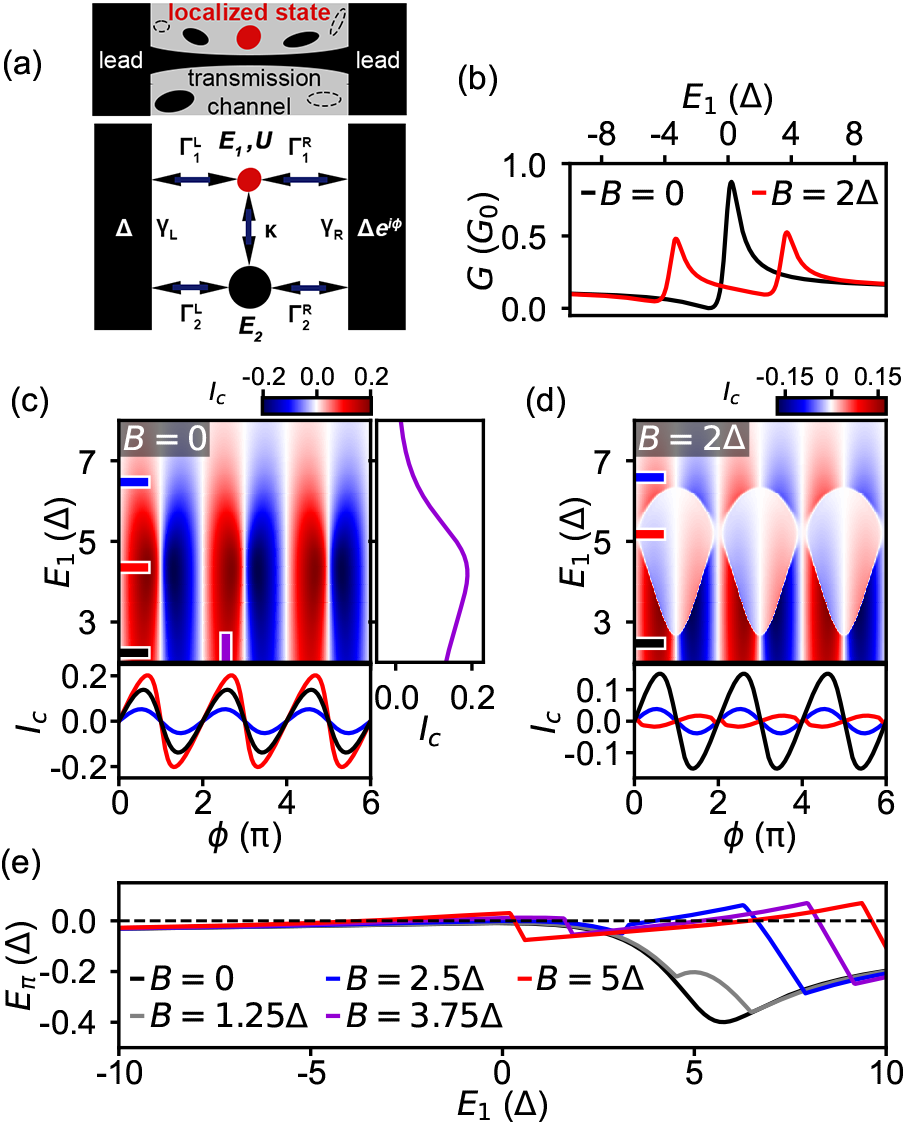}
\caption{\textbf{(a)} Schematic of a nanowire JJ and the geometry of electron distribution in the nanowire and the leads - with empty states in grey and filled states in black and red (localized state) (top). Schematic of the two-dot model - with the red dot (energy $E_1$, charging energy $U$) modelling the localized state and the black dot (energy $E_2$, no charging energy) modelling the transmission channel (bottom). The tunnel couplings to the leads (left,right) are $\Gamma_{1,2}^{L,R}$, the tunneling rate between the dots is $\kappa$ and the dots-superposition tunneling rates to the leads are $\gamma_{L,R}$.  \textbf{(b)} Normal state conductance $G$ as a function of $E_1$ for the zero (black)  and finite (red) Zeeman energy. The tunneling parameters are taken in ratios $\Gamma_1^L:\Gamma_1^R:\kappa=0.001:0.001:1$ and $\Gamma_2^L:\Gamma_2^R:E_2=0.33:0.67:1.2$. \textbf{(c)} $I_c$ in the units of $2e\Delta/\hbar$ as a function of $\phi$ and $E_1$ for $B=0$. Horizontal linecuts (blue, red and black marker) are shown below, and a vertical linecut (purple marker) is shown on the right. \textbf{(d)} Analogous to (c), but for $B=2\Delta$. \textbf{(e)} Ground-state energy difference $E_\pi=E(\phi=0)-E(\phi=\pi)$ as a function of  $E_1$ for different Zeeman energies. In (b)-(e), the charging energy is neglected by taking $U=0$.}\label{fig:pi_4}
\end{figure}

Common features in Fig. \ref{fig:pi_2} and Fig. \ref{fig:pi_3} include the peculiarly sharp and asymmetric gate voltage dependence of the supercurrent amplitude in the narrow gate intervals associated with $0-\pi$ transitions. The $\pi$-shifted CPR in our experiment occurs over $\sim 10\,\mathrm{mV}$-wide intervals of the gate voltage $V_{G1}$, which correspond to $\sim 5\Delta$-wide intervals of the junction electro-chemical potential. Sharp dependences generally suggest that a localized state is involved, and the presence of a localized state in our experiment is confirmed by detecting resonances in the normal-state conductance of the junction (Fig. \ref{fig:pi_2}(a)). The supercurrent and normal-state conductance generally remain finite when the localized state is off-resonant, which implies the existence of a background transport channel. 

The simultaneous presence of the localized state and the background transmission motivates us to explain the asymmetric resonant features by considering the interference between the localized state and direct transmission. Our motivation originates from such mechanism giving rise to the peculiar asymmetry of Fano resonances \cite{pi_fano_pr_1961}. 

We develop a model for the transport through a nanowire JJ, in which a localized state and a direct-transmission channel are involved. The top panel in Fig. \ref{fig:pi_4}(a) shows a schematic of a nanowire JJ with filled states in the leads and the transmission channel given in black. Random potential minima in the junction are either filled (small black regions) or empty (regions with dashed boundaries). The localized state is one such random minimum and is shown in red. For convenience, we treat the direct-transmission channel as a resonance as well, but its energy broadening by far exceeds all other energy scales in the model. Therefore, we develop a two-dot model of a JJ, that is introduced in the bottom panel of Fig. \ref{fig:pi_4}(a). The first (red) dot represents the localized state and the second (black) dot models the direct-transmission channel. The dot energies and tunnel couplings to the leads should, thus,  satisfy the relation $\Gamma^{L,R}_2, E_2 \gg \Gamma^{L,R}_1, E_1$ in order for the second dot to model the transmission channel. Then, we can neglect the influence of the gate voltage and magnetic field on the second dot, and we also neglect its charging energy. Importantly, there is a direct tunnel coupling with the rate $\kappa$ between the two dots, that allows for the interference between them. Additional non-trivial elements are tunneling rates $\gamma_{L,R}$ that cannot be ascribed to a certain dot, but are required to describe tunneling of a superposition state of the two dots. These parameters are at an intermediate scale, $\kappa, \gamma_{L,R} \simeq \sqrt{\Gamma_1 \Gamma_2}$. The charging energy $U$ of the first dot in general cannot be neglected. In Fig. \ref{fig:pi_4}, we present results for $U$=0 and in Fig. S8 in the Supplemental Material we display results for $U>$0 \cite{sm}. A magnetic field is introduced by the Zeeman energy in a simple form $\mathbf{B \cdot  \sigma}$, where we use $B$ to represent Zeeman energies in the units of $\Delta$. Spin-orbit coupling is neglected here, but its influence is discussed in the extended data sets in the Supplemental Material \cite{sm}. The full derivation of the model is shown in the Supplemental Material as well. 

Figure \ref{fig:pi_4}(b) shows the normal-state conductance $G$ through the two-dot system as a function of $E_1$ for $B=0$ (black) and $B=2\Delta$ (red). The ratio of the coupling rates in the model is chosen such that the total coupling to the leads is $\Gamma=4\Delta$ (as in the experiment) and the two-dot interference results in competing processes of resonant transmission and resonant reflection that almost compensate - causing the Fano shape of the resonant peculiarity \cite{pi_fano_pr_1961}. The coupling parameters remain fixed in the rest of the study \cite{sm}. Next, we perform calculations on the supercurrent transport via the two coupled dots. In Figs. \ref{fig:pi_4}(c) and \ref{fig:pi_4}(d), the junction CPR $I_c(\phi)$ is obtained as a function of $E_1$ for $B=0$ and $B=2\Delta$, respectively. For $B=0$, the $I_c$ amplitude is enhanced and asymmetrically modulated around the resonance, as confirmed by a vertical linecut. Three horizontal linecuts show that no phase shifts occur and that the CPR is skewed at the resonance, in agreement with the enhanced transmission. For $B=2\Delta$, the $I_c$ dependence exhibits three distinct regions along the $E_1$-axis - including a $\pi$-region and two $0$-regions. The $I_c$ amplitude declines inside the $\pi$-region (red linecut) and is asymmetrically modulated in the two $0$-regions (blue and black linecut). In order to more easily identify $\pi$-regions in our calculations, we define a quantity $E_\pi=E(\phi=0)-E(\phi=\pi)$ that is the difference between the junction ground state energies at $\phi=0$ and $\phi=\pi$. Therefore, a $\pi$-shifted CPR is obtained whenever $E_\pi>0$, as the ground state is favored for $\phi=\pi$. In Fig. \ref{fig:pi_4}(e), we calculate $E_\pi$ as a function of $E_1$ for different $B$. For small $B$, $E_\pi$ remains negative in the entire range of $E_1$ - confirming the absence of $\pi-$shifts at $B=0$. However, if $B$ is sufficiently large, one obtains intervals in $E_1$ with $E_\pi>0$. These intervals correspond to $\pi$-regions that appear due to the Zeeman energy - as in the example in Fig. \ref{fig:pi_4}(d). As $B$ increases, the intervals of $E_1$ with $E_\pi>0$ extend, which indicates that the $\pi$-regions broaden with the Zeeman energy.      

The theoretical model reproduces the magnetic field-driven $0-\pi$ transitions reported in the experiment. The supercurrent suppression inside the $\pi$-regions and the asymmetrical modulation outside the $\pi$-regions have been captured by the model in which the interference between the direct-transmission channel and the localized state is considered. $0-\pi$ transitions at zero magnetic field are also reproduced by the model with a sufficiently large on-site interaction and the typical features of suppressed supercurrent inside the $\pi$-regions and the asymmetrical modulation outside the $\pi$-regions still remain (see Fig. S8 \cite{sm}). In the calculations, $I_c$ jumps show up due to the Andreev-levels crossing the Fermi energy and changing the ground state parity of the junction. In the experiment, however, the parity of the junction is not controlled and the switching current measured close to parity-transitions represents an average of the two parities. Therefore, the sharp jumps in the calculated $I_c$ data are smeared-out in the measured $I_{sw}$ data. In the model, $\pi$-regions are found to occur over $\sim 10\Delta$-wide intervals in the junction electro-chemical potential - matching the scale at which they have been observed in the experiment. 

In conclusion, we report on the CPR properties of an InSb-Al nanowire JJ in high magnetic fields. The supercurrent of the device is sharply and asymmetrically modulated in narrow intervals of the junction electro-chemical potential where a localized state is involved in the transport. In these intervals, high parallel magnetic fields can drive $0-\pi$ transitions with $\pi$-shifted CPR in-between two $0$-regions. The $0-\pi$ transitions are favored by the on-site interaction in the localized state and can also occur at zero magnetic field. These phenomena can be explained by a theoretical model which involves a direct-transmission channel and a resonant localized state inside a single JJ. When one considers the interference between the direct transmission and the localized state, the supercurrent obtained in an effective Fano-resonance regime exhibits CPR features as in the experiment. Our study, thus, introduces a superconducting counterpart of the Fano effect and shows how such effect can lead to $0-\pi$ transitions in high magnetic fields.                               
\section{Acknowledgments}
We thank Ghada Badawy, Sasa Gazibegovic and Erik P. A. M. Bakkers for growing the InSb nanowires. We thank Raymond Schouten, Olaf Benningshof and J. Mensingh for valuable technical support. This work has been financially supported by the Dutch Organization for Scientific Research (NWO), Microsoft Corporation Station Q and the European Research Council (ERC) under the European Union's Horizon 2020 research and innovation program (Grant Agreement No. 694272). \\     
\bibliography{references}

\end{document}


\hsize\textwidth\columnwidth\hsize\csname@onecolumnfalse\endcsname

\title{Supplemental Material: Supercurrent in the presence of direct transmission and a resonant localized state}

\author{Vukan Levajac}
\address{QuTech and Kavli Institute of Nanoscience, Delft University of Technology, 2600 GA Delft, The Netherlands}

\author{Hristo Barakov}
\address{Kavli Institute of Nanoscience, Delft University of Technology, 2628 CJ Delft, The Netherlands}

\author{Grzegorz P. Mazur}
\author{Nick van Loo}
\author{Leo P. Kouwenhoven}
\address{QuTech and Kavli Institute of Nanoscience, Delft University of Technology, 2600 GA Delft, The Netherlands}

\author{Yuli V. Nazarov}
\email{y.v.nazarov@tudelft.nl}
\address{Kavli Institute of Nanoscience, Delft University of Technology, 2628 CJ Delft, The Netherlands}

\author{Ji-Yin Wang}
\email{wangjiyinshu@gmail.com}
\address{QuTech and Kavli Institute of Nanoscience, Delft University of Technology, 2600 GA Delft, The Netherlands}
\address{Beijing Academy of Quantum Information Sciences, 100193 Beijing, China}

\date{\today}

\pacs{}

\maketitle

\vskip1.5truecm

\section{Measurement setup}

All transport measurements are performed at $\sim 20\,\mathrm{mK}$ base temperature inside a dilution refrigerator equipped with a vector magnet.

The conductance measurement in Fig. 2(a) is performed in a two-terminal setup with standard lock-in configuration. A voltage-source sets a dc-bias voltage $V_b$ between the source and drain, and a current-meter measures a dc-current $I$ through the device. A lock-in amplifier sets an ac-bias voltage $dV_b$ which amplitude is $10\,\mathrm{\mu V}$, and measures the ac-current $dI$. Values of the bias voltages are corrected for a serial resistance $R_s=8.89\,\mathrm{k\Omega}$ as $V_b\rightarrow V_b-IR_s$ and $dV_b \rightarrow dV_b-dIR_s$. The $R_s$ includes the resistance of the voltage-source and the current-meter amplifier, the resistance of two fridge lines and the resistance of low-pass filters in the circuits.

The switching current measurements are performed in a four-terminal setup in which two terminals are connected to a current-source setting a bias current $I_b$, and the other two terminals are connected to a voltmeter measuring a voltage drop $V$ across the device. Depending on the way how $I_b$ is ramped, either a slow or a fast method for switching current measurement is used. When using the slow method, $I_b$ is swept in steps of $20-40\,\mathrm{pA}$, and $V$ is recorded for each $I_b$. Switching current can then be extracted from the recorded $V-I_b$ traces. The slow method was used for the 2D-maps in Fig. 1(b).

\begin{figure*} 
\includegraphics[width=0.95\linewidth] {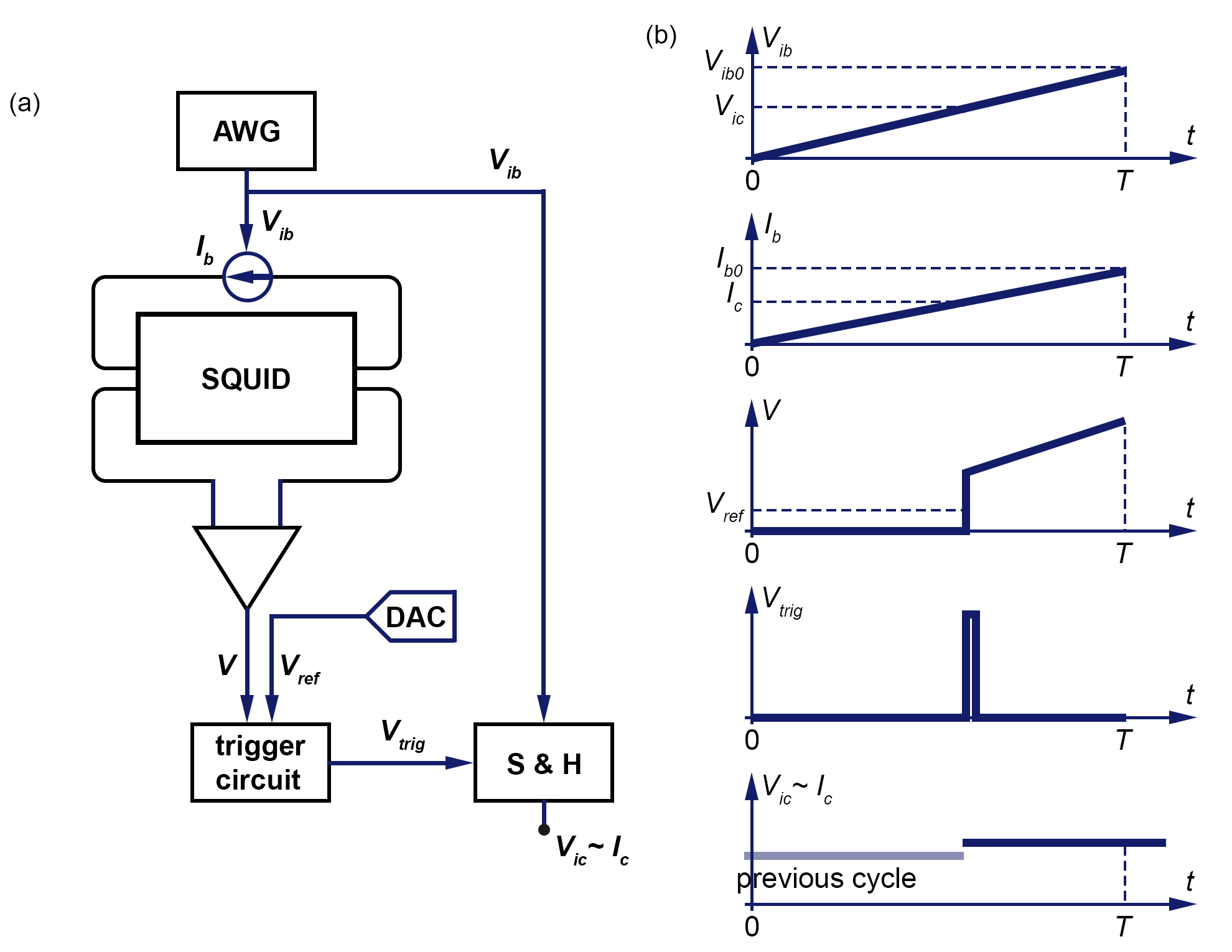}
\caption{Setup for fast switching current measurements: \textbf{(a)} schematic and \textbf{(b)} time-domain traces of the signals shown in (a). Trigger pulse $V_{trig}$ samples the voltage $V_{ic}$ that sets the bias current $I_b=I_c$ for which the switch in $V$ is detected.}
\label{fig:pi_S1}
\end{figure*}

A scheme of the setup for fast switching current measurements is shown in Fig. \ref{fig:pi_S1}(a) and time traces of relevant signals are shown in Fig. \ref{fig:pi_S1}(b) - for a single measurement period. This setup was used for collecting the data corresponding to the $I_{sw}$ data in the main text - red traces in Fig. 1(b), Fig. 2(b)-(c) and Fig. 3. The current source is controlled by an arbitrary-waveform generator (AWG) applying a sawtooth waveform of voltage $V_{ib}$ (maximal value $V_{ib0}$) at a frequency of $10\,\mathrm{Hz}$ (period $T=100\,\mathrm{ms}$). Consequently, $I_b$ is ramped with the same period in the range $[0,I_{b0}]$, where $I_{b0}$ is the maximal bias current. Here, $V_{ib0}$ is pre-selected such that $I_{b0}$ exceeds the switching current to be measured. A time-signal of the measured voltage drop $V$ and a constant voltage $V_{ref}=7\,\mathrm{\mu V}$ set by a digital-to-analog converter (DAC) are sent as inputs to a trigger circuit. As $V$ crosses the threshold $V_{ref}$, the trigger circuit sends a narrow trigger pulse $V_{trig}$ to a sample-and-hold (S\&H) circuit. The S\&H circuit receives the AWG signal as another input and this signal is sampled by each trigger pulse, and held as the output $V_{ic}$. Therefore, $V_{ic}$ represents the AWG voltage that sets the current bias for which the switch in the voltage $V$ is detected. Switching current $I_{c}$ is then extracted for a single AWG period from the conversion between the
AWG and the current-source - as $I_{c}=\frac{I_{b0}}{V_{ib}}V_{ic}$. In our work, $I_c$ was extracted for five AWG periods for each parameter set-point. The final $I_{sw}$ was calculated as an average of the five $I_c$ values. Delays due to the parasitic capacitance in the circuit, in principle, can cause an overestimate of $I_c$ - as the $V_{ic}$ output is updated with a delay with respect to the switch in $V$. However, from the specifications of the components used in our setup, this delay is estimated to be $\sim 25\,\mathrm{\mu s}$ at the given frequency and thus has practically negligible effect in our measurements where each ramp takes 100\,ms.

\section{Theoretical model and its elaboration}
 In this Section, we establish a theoretical model
to describe the normal and superconducting transport in the situation where a featureless direct transmission in a single channel is combined with that via a resonant state. Technically, it is derived from a two quantum dot model where the channel is represented as a dot with the level width that exceeds much the level width of the dot representing the resonant state. We believe that the model is applicable and useful in many situations not being restricted to concrete experimental conditions at hand. This is why we give here a detailed derivation, include the factors like strong spin-orbit interaction that are not manifested in our experiment, and provide the general examples not necessarily related to the current observations.  

\subsection{Hamiltonians}
\label{sec:ham}
In this subsection, we give the Hamiltonians of the constituents of our model.
\subsubsection{The single dot}
We start with a dot Hamiltonian. It involves on-site annihilation operators $\hat{d}_\alpha$,
$\alpha$ being the spin index, and reads
\begin{equation}
\hat{H}_D = \hat{d}^\dagger_\alpha H_{\alpha \beta}\hat{d}_\beta + U \hat{n}_\uparrow \hat{n}_\downarrow
\end{equation}
$\hat{n}_\alpha = \hat{d}^\dagger_\alpha \hat{d}_\alpha$. The single-particle Hamiltonian reads
$$
\check{H} = E + {\bf B}\cdot \boldsymbol{\sigma}
$$
${\bf B}$ being the magnetic field, $\boldsymbol{\sigma}$ being the vector of Pauli matrices.

Importantly, we treat the interaction in the mean-field approximation. If there is a natural quantization axis (that can be absent in the presence of spin-orbit interaction in the coupling to the leads), the mean field gives the following  additions to the single-particle Hamiltonian, 
\begin{equation}
H_{\uparrow\uparrow} = U \langle \hat{n}_\downarrow\rangle;  H_{\downarrow\downarrow} = U \langle \hat{n}_\uparrow\rangle.
\end{equation}
In general situation, 
\begin{equation}
H_{\alpha\beta}= U\left(\delta_{\alpha\beta}\langle \hat{N}\rangle - \langle \hat{d}^\dagger_\alpha \hat{d}_\beta\rangle \right)
\end{equation}
The advantage of this mean-field scheme is that it delivers exact results in the absence of tunnel coupling. In particular, at zero magnetic field the
ground state corresponds to single occupation of the dot in the interval
$U>E-\mu>0$. At the ends of the interval, sharp transitions bring the dot to the states of zero and double occupation. The scheme is approximate in the presence of tunnel coupling, yet we use it for the lack of better general approach to interaction. 

\subsubsection{The leads}

We introduce annihilation operators in the leads 
$\hat{c}_{k,\alpha}$ where $k$ labels the states of quasi-continuous spectrum in the leads. The states $k$ are distributed over the leads and those are labelled with $a$. We assume the states $k$ are invariant with respect to time inversion.

The leads are described by the usual BSC Hamiltonian
\begin{equation}
\hat{H}_{\rm leads} = \sum_k \xi_k \hat{c}^\dagger_{k,\alpha}\hat{c}_{k,\alpha} + \sum_a \sum_{k \in a} \left(\Delta^*_a \hat{c}_{k,\uparrow} \hat{c}_{k,\downarrow} + {\rm h.c} \right)
\end{equation}
$\xi_k$ are the energies of the corresponding states. The superconducting order parameter $\Delta_a$ is different in different leads. To describe normal leads, we just put $\Delta_a=0$.

\subsubsection{Tunnel coupling}
The tunnel coupling to the states is described by the following Hamiltonian
\begin{equation}
\hat{H}_T = \sum_k \hat{c}^\dagger_{k,\alpha} t^{k}_{\alpha\beta} \hat{d}_\beta + {\rm h.c}
\end{equation}
For time-reversible case, the tunnel amplitudes are given by
\begin{equation}
\check{t} = t_k + i {\bf t}_k \cdot \boldsymbol{\sigma}
\end{equation}
with real $t_k,{\bf t}_k$. Of course, the multitude of  tunneling amplitudes comes to the answers only in a handful of parameters. One of such parameters is the decay rate from the dot to the continuous spectrum of the lead $a$,
\begin{equation}
\Gamma_a(\epsilon) = 2\pi \sum_{k \in a} \left(|t_k|^2 + |{\bf t}_k|^2\right) \delta(\xi_k - \epsilon)
\end{equation}
One can disregard the dependence of the rates on the energy $\epsilon$.

 
\subsection{Normal transport for many dots}
\label{sec:many}
In this subsection, we will derive the currents in the nanostructure assuming the leads are normal and are kept at different filling factors. We do this derivation for an arbitrary number of the leads and dots, and later specify this for two dots and two terminals. Let us consider the following Hamiltonian where we do not specify spin or dot structure
\begin{equation}
\hat{H} = \sum_k \xi_k \hat{c}_k^\dagger c_{k} + \hat{d}^\dagger_\alpha H_{\alpha \beta}\hat{d}_\beta + \sum_k (\hat{c}^\dagger_{k} t^{k \beta } \hat{d}_\beta + h.c )
\end{equation}
The Heisenberg equations read
\begin{eqnarray}
i \dot{\hat{c}}_k = \xi_k \hat{c}_k + t_{k\alpha} \hat{d}_\alpha; \;
i \dot{\hat{d}}_\alpha = H_{\alpha\beta} \hat{d}_\beta +t^*_{k\alpha} \hat{c}_k
\end{eqnarray}
The current operators are thus given by 
\begin{equation}
\hat{I}_a = \sum_{k \in a} -i t_{k\alpha} \hat{d}^\dagger_k \hat{c}_\alpha + h.c.
\end{equation}

We solve for operators $\hat{c}_k$,

$$\hat{c}_k(t) = \hat{c}^0 e^{-i xi_k t} + \int d t' g_k(t,t') t_{k\alpha} \hat{d}_\alpha(t'),$$
$g_k(t,t') \equiv -i e^{-i\xi_k(t-t')}$, and subsequently for $\hat{d}_\alpha$,
$$
\hat{d}_\alpha(t) = \int dt' G_{\alpha\beta}(t,t') t^*_{\beta k} e^{-i \xi_k t'}\hat{d}_k^0
$$
where the Green's function obeys
\begin{equation}
\left( i \partial_t - \check{H} - \check{\Sigma}\right)\check{G} = \delta(t-t')
\end{equation}
and 
\begin{equation}
\check{\Sigma}(t,t') = \sum_k t^*_{k \alpha} g_k(t,t') t_{k \beta}.
\end{equation}
It is also useful to introduce partial $\Sigma$ that describe the decay to a certain lead,
\begin{equation}
\check{\Sigma}^a(t,t') = \sum_{k \in a} t^*_{k \alpha} g_k(t,t') t_{k \beta}.
\end{equation}
With this,
\begin{align}
\label{eq:c}
\hat{c}_k(t) = c^0_k e^{- i \xi_k t} + g_k(t,t') t_{k \alpha} G_{\alpha \beta}(t',t'') t^*_{k'\beta} e^{-i\xi_{k'} t''} \hat{c}^0_{k'} 
\end{align}
in the above expression, we assume summation over $t',t'',k'$.
We substitute this into the current operator, average over the quantum state replacing $\langle \hat{c}^{0 \dagger}_k \hat{c}^0_k\rangle = f_k$  and get two contributions corresponding to two terms in Eq. \ref{eq:c}. 
The contribution $A$ depends only on the filling factor in the lead $a$ and reads 
\begin{equation}
I^a_A = {\rm Tr} \left( \check{G}(t,t') \check{F}^a(t',t) - \check{F}^a(t,t') \check{\bar{G}}(t,t') \right)
\end{equation}
where $\bar{G}(t,t') \equiv G^\dagger(t',t)$,
\begin{equation}
\check{F}^a(t,t') = \sum_{k \in a} t^*_{k\alpha}t_{k\beta} f_k e^{-i \xi_k(t-t')}
\end{equation}
The contribution $B$ depends on filling factors in all leads
\begin{align}
I^a_B = {\rm Tr}\left( \check{G}(t,t') \sum_b \check{F}^b(t',t'') \check{\bar{G}}(t'',t''') \Sigma^\dagger_a(t''',t) - \Sigma_a(t,t') \check{G}(t',t'') \sum_b \check{F}^b(t'',t''') \check{\bar{G}}(t''',t) \right) 
\end{align}
We switch to the energy representation. To deal with the tunnel amplitudes, we will use the following relation
\begin{equation}
\label{eq:Gammadefinition}
\check{\Gamma}^a(\epsilon) = 2\pi \sum_k t^*_{k\alpha} t_{k \beta} \delta(\epsilon - \xi_k)
\end{equation}
$\check{\Gamma}^a$ characterizing the decay from all dots to the lead $a$. Conventionally, we will disregard the energy dependence of $\Gamma$ (since we are working close to the Fermi level). With this,
\begin{equation}
\check{F}^a = -i \check{\Gamma}^a f_a(\epsilon); \; \check{\Sigma}_a = - \frac{i}{2} \check{\Gamma}^a,
\end{equation}
where we have taken into account that the filling factor depends on energy only, and disregarded real part of $\Sigma$ (that would lead to a renormalization of the dot Hamiltonian). With this, the Green function is given by
\begin{equation}
\check{G} = \frac{1}{ \epsilon - \check{H} + i \check{\Gamma}/2};
\end{equation}
$\check{\Gamma} \equiv \sum_a \check{\Gamma}_a$.
The $B$ contibution for the current for all $b \ne a$
can be written as
\begin{equation}
I_a/e = \sum_{b \ne a} \int \frac{d \epsilon}{2\pi} P_{a b} (\epsilon) f_b(\epsilon)
\end{equation}
$P_{ab}$ being the  probability to scatter from all channels of terminal $b$ to the channels of terminal $a$,
\begin{equation}
P_{ab}(\epsilon) = {\rm Tr} \{\check{\Gamma}^a \check{G}(\epsilon) \check{\Gamma}^b \check{\bar{G}}(\epsilon) \}
\end{equation}
This is in accordance with the corresponding part of Landauer formula for multi-terminal case. The contibution $A$ reads:
\begin{equation}
I^a_A/e = -i \int \frac{d \epsilon}{2\pi} f_a(\epsilon) {\rm Tr} \{\check{\Gamma}^a (\check{G} -\check{\bar{G}})\}
\end{equation}
We use the relation
\begin{equation}
\label{eq:diffofgreens}
\check{G}  -\check{\bar{G}} = -i \check{G} \check{\Gamma} \check{\bar{G}}
\end{equation}
to represent the contribution $A$ in the form
\begin{equation}
I^a_A/e = - \int \frac{d \epsilon}{2\pi} f_a(\epsilon) \sum_b P_{ab}(\epsilon)
\end{equation}
summing everything together, we reproduce the Landauer formula
\begin{equation}
I_a/e = \int \frac{d \epsilon}{2\pi} \sum_{b \ne a} P_{ab}(\epsilon)(f_a(\epsilon)-f_b(\epsilon))
\end{equation}
Let us construct a scattering matrix corresponding to the situation. The scattering to a terminal $a$ is described by $\check{\Gamma}^a$. Let us represent this matrix as $\check{\Gamma}^{a} = \check{W}_a^\dagger \check{W}_a$. So-introduced $\check{W}_a$ is a matrix where the second index goes over the dots and the first one over the channels of the terminal $a$. The matrix $\check{W}_a$ is apparently an ambiguous representation of $\check{\Gamma}^a$, but the same ambiguity pertains the scattering matrix: both are defined upon a unitary transformation in the space of the channels in each lead. We combine all matrices $W_a$ block by block introducing the matrix $W$ where the first index goes over all channels in all terminals. We note $\check{W}^{\dagger} \check{W} = \check{\Gamma}$. With this, a scattering matrix describing the situation reads
\begin{equation}
\check{S} = 1 - i \check{W} \check{G} \check{W}^\dagger
\end{equation} 
Its unitarity can be proven with using the relation (\ref{eq:diffofgreens}).

\subsection{Normal transport for two dots}
\label{sec:twodots}
We concentrate on the case of two dots and two terminals. It seems a trivial consequence of the above fomulas but requires some elaboration for the limit where $ \Gamma$ in the dots are very different, this is the case under consideration. To warm up, let us first consider a single dot. We note that $\Gamma_a$ in this case are diagonal in spin owing to time-reversability and can be regarded as numbers. The transmission probability from the left to the right (or vice versa) 
can be written as 
\begin{equation}
T_0(\epsilon) = \frac{\Gamma_L \Gamma_R}{(\epsilon-E)^2 +\Gamma^2/4} 
\end{equation}
The ideal transmission is achieved at $\Gamma_L=\Gamma_R = \Gamma/2$ and $\epsilon = E$.
Let us go for two dots and list possible parameters of the model. Those are: level energies (split in spin) $E_1 + \boldsymbol{B}_1\cdot \boldsymbol{\sigma}$, $E_2 + \boldsymbol{B}_2\cdot \boldsymbol{\sigma}$, decays from the dots $\Gamma_1 = \Gamma^L_1 +\Gamma^R_1$, $\Gamma_2 = \Gamma^L_2 +\Gamma^R_2$, tunneling between the dots $\kappa + i \boldsymbol{\kappa} \cdot \boldsymbol{\sigma}$, and non-diagonal tunneling to the leads $\Gamma_{12,21} \equiv \gamma \pm i \boldsymbol{\gamma} \cdot \boldsymbol{\sigma}$. 
Let us write down the Green's function:
\begin{align}\label{eq:twodotmodel}
\check{G}^{-1} = \epsilon 
- \begin{bmatrix}
H_1  &  H_{12} \\ 
H^\dagger_{12} & H_2
\end{bmatrix} ; \;
H_{1,2} \equiv E_{1,2} + \boldsymbol{B}_{1,2}\cdot \boldsymbol{\sigma} -i\Gamma_{1,2}/2 ; \;
H_{12} \equiv \kappa + i \boldsymbol{\kappa} \cdot \boldsymbol{\sigma} -i(\gamma +i \boldsymbol{\gamma} \cdot \boldsymbol{\sigma})/2
\end{align}
The idea of further transform is that the second dot provides a featureless background for the first dot. To this end, we consider big $E_2,\Gamma_2 \gg \epsilon, B_{1,2}, E_1, \Gamma_1$.
As to $\gamma,\kappa$, they are assumed to be of an intermediate scale, say $\gamma \simeq \sqrt{\Gamma_1 \Gamma_2}$. We note that $\boldsymbol{B}_{2}$ can be ignored under this condition, and for brevity we define $\boldsymbol{B}_{1} = \boldsymbol{B}$.

We will apply a transform that approximately diagonalises the Green function so that 
\begin{equation}
\check{G} = \check{U} \check{G}_d \check{U}^{-1}
\end{equation}
where 
\begin{align}
\check{U} =\sqrt{\frac{1+s}{2s}} \begin{bmatrix}1 & \eta_+\\
-\eta_- & 1\end{bmatrix};\; 
\check{U}^{-1} =\sqrt{\frac{1+s}{2s}} \begin{bmatrix}1 & -\eta_+\\
\eta_- & 1\end{bmatrix}
\end{align}
and
\begin{align}
\eta_{\pm} = \frac{\mu_{\pm}}{1+s};\; s\equiv\sqrt{1+ \mu_+\mu_-};\;
 \mu_{\pm} = 2 \frac{k \pm \boldsymbol{k}\cdot \boldsymbol{\sigma}  }{-E_2 + i\Gamma_2/2 };\;
 k,\boldsymbol{k} \equiv -\kappa+i\gamma/2, - \boldsymbol{\kappa} + i \boldsymbol{\gamma}/2
\end{align}
with this, the biggest block of $\check{G}^{-1}_d$ is $-E_2 + i\Gamma_2/2$, while the smallest one reads
\begin{equation}
\epsilon - E_1 + i\Gamma_1/2 - \frac{k^2+\boldsymbol{k}^2}{-E_2 + i \Gamma_2/2}
\end{equation}
We rewrite it as
\begin{equation}
\epsilon - E_1 + i\Gamma/2 -\Delta E_1
\end{equation}
where the actual level width $\Gamma$ is given by 
\begin{align}
\Gamma = \Gamma_1 + \frac{\Gamma_2 C_{11} - 2 E_2 C_{10}}{E_2^2 + \Gamma_2^2/4}; \;
 C_{11} \equiv \kappa^2 - \gamma^2/4 + \boldsymbol{\kappa}^2 - \boldsymbol{\gamma}^2/4; \; C_{10} \equiv \kappa\gamma + \boldsymbol{\kappa}\boldsymbol{\gamma}
\end{align} 
and we neglect insignificant shift of the level position 
\begin{equation}
\Delta E_1 = -\frac{C_{10}\Gamma_2/2 + C_{11} E_2}{E_2^2 + \Gamma_2^2/4}
\end{equation} 
The $\Gamma_a$ matrices are transformed as
$\check{\Gamma}^L \to \check{U}^\dagger \check{\Gamma}^L \check{U}$, $\check{\Gamma}^L \to \check{U}^{-1\dagger} \check{\Gamma}^L \check{U}^{-1}$. 

Keeping terms of the relevant orders only, we obtain
\begin{align}
\check{\Gamma}^L = \begin{bmatrix}
g_L& \Gamma_{12}^{+L} - \eta_-^* \Gamma_2^L \\
\Gamma_{12}^{-L} - \Gamma_2^L \eta_- & \Gamma_2^L 
\end{bmatrix};\;
g_L \equiv \Gamma_1^L - \Gamma_{12}^{+L} \eta_- -\eta_-^* \Gamma_{12}^{-L} + \eta_-^*\Gamma_2^L \eta_-;
\end{align}

\begin{align}
\check{\Gamma}^R = \begin{bmatrix}
g_R & \Gamma_{12}^{+R} - \eta_+ \Gamma_2^R \\
\Gamma_{12}^{-R} - \Gamma_2^R \eta_+^* & \Gamma_2^R 
\end{bmatrix} ; \;
g_R \equiv \Gamma_1^R - \Gamma_{12}^{+R} \eta_+^* -\eta_+ \Gamma_{12}^{-R} + \eta_+\Gamma_2^R \eta_+^*-. 
\end{align}

With this, we can  summarize the results for the total transmission coefficient $T_{tot}$ (summed over two spin directions). We introduce compact notations that adsorb the energy dependence of the coefficient:
\begin{align}
G_{\pm} = \frac{1}{\epsilon - E_1 \pm B +i\Gamma/2};\; G_{s,a} = \frac{G_+\pm G_-}{2};\ \bar{G}_i = G^*_i
\end{align}
and write it down as
\begin{eqnarray}
T_{tot}(E)=2T_0 
+(\Gamma_L \Gamma_R +\boldsymbol{\Gamma}^2) (G_+\bar{G}_+ + G_-\bar{G}_-) 
+ 2((\boldsymbol{\Gamma}\cdot \boldsymbol{B})^2/B^2 -\boldsymbol{\Gamma}^2) G_a\bar{G}_a \\
+ RX (G_++G_-+\bar{G}_++\bar{G}_-) 
- IX {\rm Im} (G_++G_--\bar{G}_+-\bar{G}_-) \label{eq:totalT}
\end{eqnarray}
Here, the partial decay rate read ($\Gamma_L+\Gamma_R =\Gamma$)
\begin{eqnarray}
\Gamma_L = \Gamma^L_1 +\frac{C_1 \Gamma^L_2-C^L_3 \Gamma_2 - 2 E_2 C_2^L}{E_2^2 +\Gamma_2^2/4};\;
C_1 \equiv \kappa^2 + \gamma^2/4 + \boldsymbol{\kappa}^2 + \boldsymbol{\gamma}^2/4;\;
C_2^L \equiv \boldsymbol{\kappa}\cdot \boldsymbol{\gamma}_L + \gamma_L \kappa;\;
C_3^L \equiv \boldsymbol{\gamma}\cdot \boldsymbol{\gamma}_L +\gamma \gamma_L,
\end{eqnarray}
and similar for $R$.
The spin-orbit interaction is represented by the vector $\boldsymbol{\Gamma}$,
\begin{eqnarray}
\boldsymbol{\Gamma} &=& \frac{E_2 \boldsymbol{C}_5 +\boldsymbol{\kappa} C_4 + \boldsymbol{C}_6\times \boldsymbol{\kappa} +\kappa \boldsymbol{C}_6}{E_2^2 +\Gamma_2^2/4};\;
C_4 = \Gamma_2^L \gamma_R - \Gamma_2^R \gamma_L ;\\
\boldsymbol{C}_5 &=& \boldsymbol{\gamma}_R \gamma_L - \boldsymbol{\gamma}_L \gamma_R + \boldsymbol{\gamma}_R \times \boldsymbol{\gamma}_L ;\;
\boldsymbol{C}_6 = \Gamma_2^R \boldsymbol{\gamma}_L -\Gamma_2^L \boldsymbol{\gamma}_R
\end{eqnarray}
and the coefficients $RX$, $IY$ read
\begin{align}
RX &= \frac{1}{E_2^2 +\Gamma_2^2/4}(-E_2 C_7 +\kappa C_8 + \boldsymbol{\kappa} \cdot \boldsymbol{C}_9 
- T_0 (E_2 C_{11} + C_{10} \Gamma_2/2))&\label{eq:RX}\\
IX &= \frac{1}{E_2^2 +\Gamma_2^2/4} (- C_7\Gamma_2/2 +\gamma C_8/2 +
\boldsymbol{\gamma} \cdot \boldsymbol{C}_9/2 
 -T_0 (E_2 C_{10} - C_{11} \Gamma_2/2)) &\\
C_7 &= \gamma_R \gamma_L + \boldsymbol{\gamma}_R \cdot\boldsymbol{\gamma}_L ;\; 
C_8 = \Gamma_2^L \gamma_R + \Gamma_2^R \gamma_L;\;
\boldsymbol{C}_9 =  \boldsymbol{\gamma}_R \Gamma_2^L + \boldsymbol{\gamma}_L \Gamma_2^R &
\end{align}
We will explain the physical significance of each term in Eq. \ref{eq:totalT} in the next subsection.

To treat the interaction self-consistently, we also need the average charge and spin in the dot,
\begin{equation}
\langle \hat{d}^\dagger_\alpha \hat{d}_\beta \rangle \equiv n \delta_{\alpha\beta} + \boldsymbol{n} \cdot \boldsymbol{\sigma} 
\end{equation}
This is given by 
\begin{align}
\check{n} = \int \frac{d\epsilon}{2\pi} \check{G}\left( (\Gamma_R + \boldsymbol{\Gamma}\cdot \boldsymbol{\sigma})f^R(\epsilon)+(\Gamma_L - \boldsymbol{\Gamma}\cdot \boldsymbol{\sigma})f^L(\epsilon) \right) 
\end{align} 
This can be rewritten in more detail as ($\boldsymbol{b} =\boldsymbol{B}/B$)
\begin{eqnarray}
n &=& \int \frac{d\epsilon}{2\pi} \left((G_s \bar{G}_s + G_a \bar{G}_a) (\Gamma_R f^R(\epsilon)  
 + \Gamma_L f^L(\epsilon))  + (\boldsymbol{b} \cdot \boldsymbol{\Gamma}) (G_a \bar{G}_s + G_s \bar{G}_a) (f^R(\epsilon)-f^L(\epsilon)) \right) \\
\boldsymbol{n} &=& \int \frac{d\epsilon}{2\pi} \left(2 \boldsymbol{b} (\boldsymbol{b}\cdot \boldsymbol{\Gamma}) G_a \bar{G}_a +\boldsymbol{\Gamma} (G_s \bar{G}_s - G_a \bar{G}_a) + 
(\boldsymbol{b} \times \boldsymbol{\Gamma}) i (G_a \bar{G}_s - G_s \bar{G}_a)) (f^R(\epsilon)-f^L(\epsilon)) \right. \nonumber \\ &+& \left.
\boldsymbol{b} (G_a \bar{G}_s + G_s \bar{G}_a) (\Gamma_R f^R(\epsilon) + \Gamma_R f^R(\epsilon) \right)
\end{eqnarray}

We substitute filling factors at vanishing temperature $f^{L,R} = \Theta(eV_{L,R} - \epsilon)$ and integrate over $\epsilon$ to obtain $n, \boldsymbol{n}$ and full current. It is also advantageous at this stage to switch to dimensionless variables measuring energy in units of $\Gamma$ and setting $e=1$. We introduce convenient functions
\begin{eqnarray}
K^{\pm}_{R,L} = \frac{1}{2\pi} {\rm atan}(2(V_{R,L}-\epsilon_d \pm B));\;
L^{\pm}_{R,L} = \frac{1}{2\pi} {\rm ln}( 4(V_{R,L} \pm B)^2+1);\;
L^{\pm} = L^{\pm}_{R} - L^{\pm}_{L}; \; K^{\pm} = K^{\pm}_{R} - K^{\pm}_{L}.
\end{eqnarray}
With this, 
\begin{eqnarray}
n&=& \sum_{k=L,R} \Gamma_k (1/2 +K^+_k +K^-_k)  
+(\boldsymbol{b} \cdot \boldsymbol{\Gamma}) (K^++K^-), \\
\boldsymbol{n} &=& \boldsymbol{b} \left( \Gamma_R (K^-_R -K^+_R) + \Gamma_L (K^-_L -K^+_L)\right) 
+ \frac{\boldsymbol{\Gamma} }{1+4 B^2} \left(K^++K^- +B(L^--L^+)\right) \nonumber \\
&+& \frac{(\boldsymbol{b} \times \boldsymbol{\Gamma}) }{2(1+4 B^2)}  \left(B(K^++K^-) +L^+-L^-\right) 
+ \frac{2 \boldsymbol{b} (\boldsymbol{b} \cdot \boldsymbol{\Gamma}) B}{1+4 B^2}\left(4B(K^++K^-) +L^+-L^-\right) 
\end{eqnarray}
The self-consistency equations then read:
\begin{equation}
\epsilon_d = U n; \; \boldsymbol{B} = \boldsymbol{B}_0 - U \boldsymbol{n}
\end{equation}
$\boldsymbol{B}_0$ being the external magnetic field. This equation has to be solved at each $V_{R,L}$. With this solution, we can evaluate the current 
\begin{eqnarray}
I &=&  T_0(V_L-V_R)/\pi + 2 (\Gamma_L\Gamma_R +\boldsymbol{\Gamma}^2) (K^++K^-) 
+ 4 ((\boldsymbol{\Gamma}\cdot \boldsymbol{b})^2 - \boldsymbol{\Gamma}^2)\frac{B}{1+4 B^2} \left(4B(K^++K^-) +L^+-L^-\right) \nonumber \\
&+& RX (L^++L^-) - IX (K^++K^-)/2 \label{eq:currentself}
\end{eqnarray} 

Let us elaborate on the equilibrium case $V_R=V_L=\mu$. In this case, owing to time-reversibility, the spin-orbit interaction does not cause spin polarization and is irrelevant, so the self-consistency equations read ($\tilde{K}=K^R=K^L$)
\begin{align}
\epsilon_d = U (1/2 + \tilde{K}^+ +\tilde{K}^-); 
 \boldsymbol{B} =\boldsymbol{B}_0 -\boldsymbol{b} U (\tilde{K}_- - \tilde{K}_+)
\end{align}
We specify to $\boldsymbol{B}_0 =0$ and determine the boundary of spontaneously magnetic phase where $B \to 0$. In this limit,
\begin{equation}
\tilde{K}_- - \tilde{K}_+ \to  - B\frac{2}{\pi} \frac{1}{1+4(\mu^*)^2}; \; \mu^* = \mu - \epsilon_d
\end{equation}
with this, the equations for the boundary read
\begin{align}
U= (1+4(\mu^*)^2)\frac{\pi}{2}; \;
 \mu = \mu^* + U (1/2 + (1/\pi) {\rm arctan}(2 \mu^*))
\end{align}
The splitting occurs above critical value $U_c = \pi/2$, at large $U$ the magnetic phase occurs in the interval $\mu = (0,U)$ as it should be in this limit.

\subsection{ Normal transport examples}
\label{sec:normaltransportexamples}
In this subsection, we will analyse the peculiarities of normal transport in the model at hand. We restrict ourselves to zero-voltage conductance and non-interacting case where zero-voltage conductance is simply given by $T_{tot}$ at $\epsilon$ corresponding to Fermi level,
\begin{equation}
G(V_g) = \frac{G_Q}{2} T_{tot}(\epsilon=E_F).
\end{equation}
Since $E_1$ is a linear function of the gate voltage, and shift of $\epsilon$ in Eq. \ref{eq:totalT} is equivalent to the shift of $E_1$, the energy dependence of $T_{tot}$ directly gives the gate voltage dependence of the conductance. The conductance with interaction is qualitatively similar to the non-interacting one since the main effect of interaction in our model is the spin-splitting corresponding to $B \simeq U$.

Let us explain the physical significance of the terms in Eq. \ref{eq:totalT}. All spin-orbit effects are incorporated into a single vector $\boldsymbol{\Gamma}$ in the spin space. To start with, let us neglect the spin-orbit interaction setting $\boldsymbol{\Gamma}=0$, so we can disregard the third term. In this case, $T_{tot}$ is contributed independently by spin orientations $\pm$ with respect to $\boldsymbol{B}$. Their contributions are shifted by $2B$ in energy.

The first term in Eq. \ref{eq:totalT} gives the featureless transmission of the transport channel and asymptotic value of the conductance at $|E_1| \gg \Gamma$. The second term describes the resonant transmission via the localized state and would show up even if there is no interference between the transmissions through the channel and the localized state. It rives rise to a Lorentzian peak - resonant transmission - of the width $\simeq \Gamma$ in conductance that splits into two at sufficiently big spin splitting $\simeq \Gamma$. Let us bring the fifth term into consideration. Since $G-\bar{G} = -i \Gamma G\bar{G}$ its energy dependence is identical to the second one. However, it usually gives a negative contribution to transmission describing destructive interference of the transmissions in the dot and in the channel - resonant reflection. 

The fourth term describes the celebrated Fano effect coming about the interference of the resonant and featureless transmission. It is visually manifested as asymmetry of otherwise Lorentzian peaks or dips. The antisymmetric Fano tail $\propto \epsilon^{-1}$ at large distances from the peak/dip centre beats Lorentzian tail $\propto \epsilon^{-2}$. All these terms are hardly affected by spin-orbit interaction, while the second one manifests it fully. It mixes up spin channels and makes conductance to depend on the orientation of $\boldsymbol{B}$ with respect to $\boldsymbol{\Gamma}$. 

\begin{figure*}
\begin{center}
	\includegraphics[width=\textwidth]{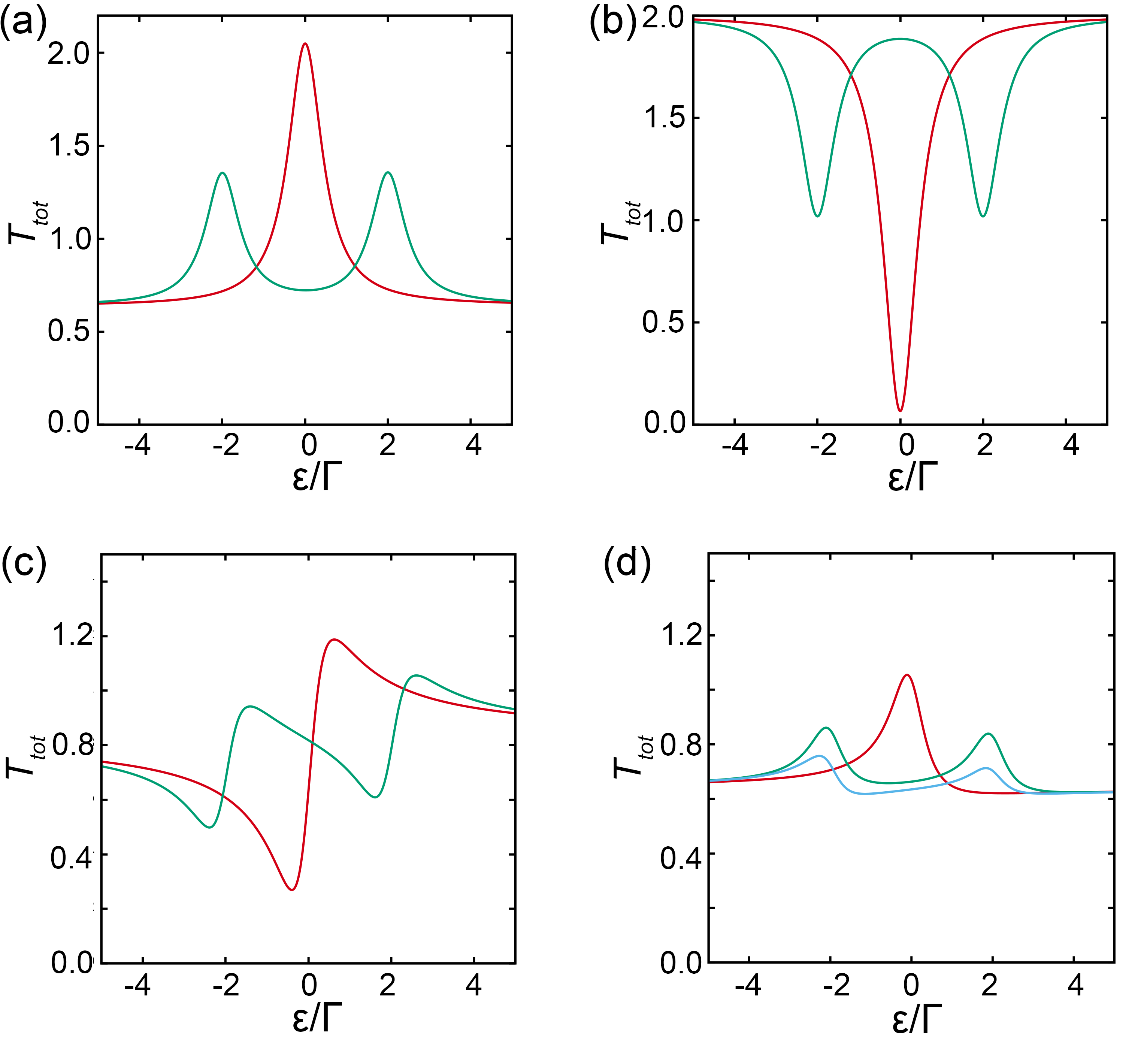}
\end{center}
\caption{Examples of normal transport. The energy dependence of $T_{tot}$ is the same as the conductance dependence on the gate voltage. Red curves correspond to $B=0$, green curves to $B=2\Gamma$.  a. Basic example: resonant transmission b. Dip: resonant reflection c. Fano. d. Strong spin-orbit. Here, green (blue) curve is for parallel (perpendicular) orientation of $\boldsymbol{B}$ with respect to $\boldsymbol{\Gamma}$. }
\label{fig:normaltransport}
\end{figure*}

We illustrate the possible forms of the energy (or, equivalently, gate-voltage) dependence of the conductance with the plots in Fig. \ref{fig:normaltransport} for 4 settings of the parameters $\Gamma_2^{L,R}, E_2, \kappa, \boldsymbol{\kappa}, \gamma_{L,R}, \boldsymbol{\gamma}_{L,R}$. Owing to separation of the scales assumed, the relevant parameters $\Gamma_{L,R}, \boldsymbol{\Gamma}, RX, IX$ are invariant with respect to rescale with the factor $A$,
\begin{align}
\Gamma_2^{L,R}, E_2 \to A (\Gamma_2^{L,R}, E_2) ;\;
\kappa, \boldsymbol{\kappa}, \gamma_{L,R}, \boldsymbol{\gamma}_{L,R} \to \sqrt{A} (\kappa, \boldsymbol{\kappa}, \gamma_{L,R}, \boldsymbol{\gamma}_{L,R}).
\end{align}
For all settings, energy is in units of the resulting $\Gamma$. For each setting, we give the plots at $B=0$ and $B = 2 \Gamma$, the latter to achieve a visible separation of resonant peculiarities. Spin-orbit interaction is weak except the last setting where we give separate plots for $\boldsymbol{B} \parallel \boldsymbol{\Gamma}$ and $\boldsymbol{B} \perp \boldsymbol{\Gamma}$. 

For Fig. \ref{fig:normaltransport} (a) we choose $\Gamma_2^L,\Gamma_2^R,E_2 = A(0.2,0.8,0.5)$, $\kappa, \gamma_L,\gamma_R = \sqrt{A}(0.5,0.2,0.2)$, $\Gamma_1^L, \Gamma_1^R = 1.6, 3.5$. We also specify small but finite spin-orbit terms yet they hardly affect the conductance. In this case, the transmission through the localized state is faster than the interference with the transmission in the channel. This results in a resonant reflection peak at $B=0$ that splits into two upon increasing the magnetic field. A little Fano asymmetry can be noticed upon a close look.

For Fig. \ref{fig:normaltransport} (b) we choose $\Gamma_2^L,\Gamma_2^R,E_2 = A(0.5,0.5,0)$, $\kappa, \gamma_L,\gamma_R = \sqrt{A}(3.5,0.2,0.2)$, $\Gamma_1^L, \Gamma_1^R = 0.5, 0.5$.
The transmission through the channel is ideal, $T_0 =1$. The localized state is connected to the channel better than to the leads ($\kappa \gg \gamma_{L,R}$). This results in a pronounced resonant reflection dip at $B=0$ that also splits into two upon increasing the magnetic field.

For Fig. \ref{fig:normaltransport} (c) we choose $\Gamma_2^L,\Gamma_2^R,E_2 = A(0.2,1.5,0)$, $\kappa, \gamma_L,\gamma_R = \sqrt{A}(1.5,0.3,0.1)$, $\Gamma_1^L, \Gamma_1^R = 0.8, 0.1$. This choice is such that the competing processes of resonant transmission and reflection almost compensate each other so the resulting resonance peculiarity assumes almost antisymmetric Fano shape. The separation of the peculiarities upon the spin splitting is less pronounced than in the previous examples owing to long-range Fano tails mentioned.

We illustrate the effect of strong spin-orbit interaction in Fig. \ref{fig:normaltransport} (d).  We choose $\Gamma_2^L,\Gamma_2^R,E_2 = A(0.2,0.8,0.5)$, $\kappa, \gamma_L,\gamma_R = \sqrt{A}(0.5,0.2,0.2)$, $\Gamma_1^L, \Gamma_1^R = 1.6, 3.5$. 
As to spin-dependent parameters, we choose 
$\boldsymbol{\kappa} = \sqrt{A} SO [0, 0.2, -0.6]$, $\boldsymbol{\gamma}_L = \sqrt{A} SO [0.3,0,0]$, $\boldsymbol{\gamma}_R = \sqrt{A} SO [0.0,0,1]$ and set the coefficient $SO$ to $1.6$, this is its maximal value that satisfies the positivity conditions imposed on the matrices of the rates. The peculiarity at $B=0$ is a peak with a noticeable Fano addition. It splits at $B= 2 \Gamma$ changing its shape, that is different for $\boldsymbol{B} \parallel \boldsymbol{\Gamma}$ and 
$\boldsymbol{B} \perp \boldsymbol{\Gamma}$ as well as for positive and negative energies. Note that owing to Onsager symmetry $
G(\boldsymbol{B}) = G(-\boldsymbol{B})$.

\begin{figure*}
\begin{center}
	\includegraphics[width=\textwidth]{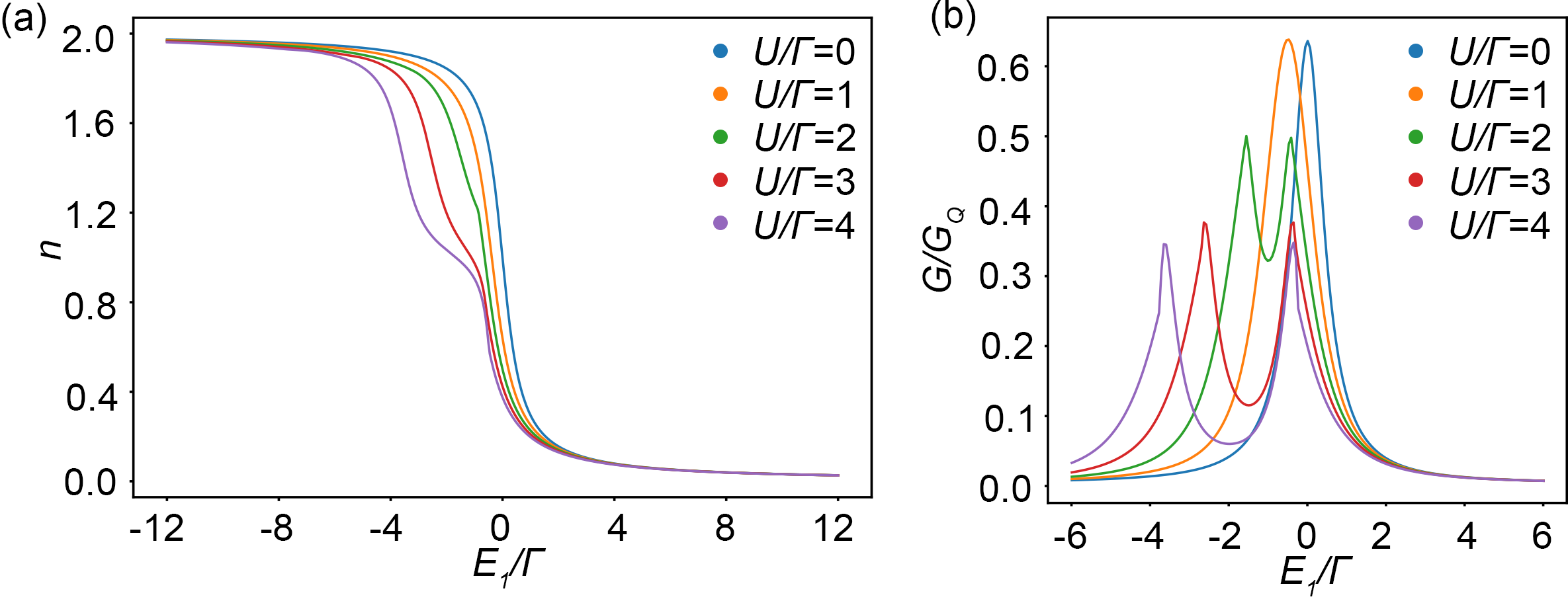}
\end{center}
\caption{Example of normal transport with interaction. Resonant transmission regime, no magnetic field, no SO coupling. The setup parameters are given in the text.  a. The average number of electrons in the localized state versus $E_1$ at various interaction strengths. b. Zero-voltage conductance versus $E_1$ at various interaction strengths.}
\label{fig:self}
\end{figure*}

We also provide an example with interaction implementing the self-consistent scheme described in the previous subsection (Fig. \ref{fig:self}). For this example, we choose $\Gamma_2^L,\Gamma_2^R,E_2 = A(0.2,1.5,-15)$, $\kappa, \gamma_L,\gamma_R = \sqrt{A}(0.8,0.1,0.1)$, $\Gamma_1^L, \Gamma_1^R = 1.1, 0.9$. This choice corresponds to very low channel transmission ($T_0 = 10^{-3}$).
The average number of electrons in the dot is presented in Fig. \ref{fig:self}(a) as a function of $E_1$ for several interaction strengths, at zero voltage difference and magnetic field. All curves change from full occupation at big negative $E_1$ to zero occupation at big positive $E_1$. At $U=0$ and $U=\Gamma$ the curves are smooth with no 
spontaneous spin splitting emerging throught the whole interval of $E_1$. For higher interaction strengths, there is an interval of $E_1$ where the spontaneous splitting is present. The ends of this interval are in principle manifested by cusps in the curves. Only cusps at the end of the interval close to zero are visible, the cusps at the other end are too small.
It might seem that the zero-voltage conductance (Fig. \ref{fig:self} (b)) can be computed from $T_{tot}$ at the parameters  $\tilde{E}_1, \tilde{B}$ that solve the self-consistency equation at zero voltage difference. However, this is not so, since these parameters also depend on voltage difference. We compute  zero-voltage conductance by numerically differentiating the current (Eq. \ref{eq:currentself}) at small voltage differences. At zero interaction, we see a resonant transmission peak. Its height does not reach $G_Q$ because of the asymmetry $\Gamma_R \ne \Gamma_L$. At $U=\Gamma$, there is still a single peak. At higher $U$ we see the splitting of the peak. The height of the peaks split is a half of the height of the original peak if they are sufficiently separated. As we have conjectured earlier, this is qualitatively similar to the conductance traces where spin splitting is induced by the magnetic field. 

\subsection{Superconducting transport}
\label{sec:suptrans}
In this subsection, we elaborate on the description of superconducting transport in our model. Since supercurrent is a property of the ground state of the system, it is convenient to work with electron Green functions in imaginary time and introduce Nambu structure. Let us start, as we did previously, with an arbitrary number of dots and superconducting leads. If we neglect tunnel couplings, the inverse Green function ${\cal H}(\epsilon)$ is a $8 \times 8$ matrix encompassing the spin index, Nambu index and that labelling the dots. It reads:
\begin{equation}
\check{{\cal H}} = i \epsilon \tau_z - \check{H}.
\end{equation}
The tunnel couplings to the leads labelled by $a$ add the self-energy part
\begin{align}
\check{{\cal H}} = i \epsilon \tau_z - \check{H} + \frac{i}{2} \sum_a \check{\Gamma}_a \check{Q}_a
\end{align}
where $\check{\Gamma}_a$ are given by Eq. \ref{eq:Gammadefinition} and the matrix $\check{Q}_a$ is a matrix in Nambu space reflecting the properties of the superconducting lead $a$,
\begin{equation}
Q_{a}=\frac{1}{\sqrt{\epsilon^2 +\Delta_a^2}}
\begin{bmatrix}
\epsilon& \Delta_a e^{i\phi_{a}}\\
\Delta_a e^{-i\phi_{a}}& -\epsilon
\end{bmatrix},
\end{equation}
$Q_a^2=1$.

To find supercurrents, we need to evaluate the total energy
and take its derivatives with respect to the phase differences. Since the leads with different phases are connected by the dots, the phase-dependent energy is the energy of the dots. The latter can be expressed as

\begin{equation}
	{\cal E} =-\frac{1}{2}\int \frac{d\epsilon}{2 \pi} \ln \det(\check{{\cal H}})
\end{equation}

To see how this works, let us check this formula neglecting tunnel couplings.
With this, the energy is the sum over eigenvalues of $\check{H}$, $E_n$,
\begin{equation}
	{\cal E} =-\frac{1}{2}\int \frac{d\epsilon}{2 \pi} \ln (\epsilon^2 + E_n^2)
\end{equation}
The integral formally diverges at $\epsilon \to \infty$. To regularize it, we 
subtract its value at $E_n=0$ to obtain
\begin{equation}
{\cal E} = - \sum_n \frac{|E_n|}{2} + const
\end{equation}
To recover a familiar formula, we shift the constant by ${\rm Tr}(\check{H})/2$,
 \begin{align}
{\cal E} = - \sum_n \frac{|E_n|}{2} + \sum_n \frac{E_n}{2} + const =
\sum_n E_n \Theta(-E_n) + const, 
\end{align}
so it becomes the energy of the filled states (those with $E_n<0)$.
This suggests that we need to handle the integral with care keeping eye on possible problems at big $\epsilon$.
Fortunately, no special care has to be taken for the phase-dependent energy since it is accumulated at superconducting gap scale $\epsilon \simeq \Delta$. We have to be careful when expressing the occupation of the dots in terms of derivatives of ${\cal E}$ with respect to dot energies (as we do for numerical calculations). For instance, the average occupation of the dot $1$ reads
\begin{equation}
\langle \hat{n}_1 \rangle = \frac{\partial {\cal E}}{\partial E_1} + 1,
\end{equation} 
the last term correcting for high-energy divergences.

For our starting two-dot, two-lead model, the inverse Green function reads (c.f. with Eq. \ref{eq:twodotmodel}). 

\begin{equation}
{\cal H}=\begin{bmatrix}
{\cal H}_{11}&{\cal H}_{12}\\
{\cal H}_{21}&{\cal H}_{22}
\end{bmatrix}
\end{equation}
, where 

\begin{align}\label{eq:H11}
{\cal H}_{11}&=i\epsilon \tau_z-E_1  - (\boldsymbol{B}_1 \cdot \check{\boldsymbol{\sigma}})\tau_z 
+ \frac{i}{2} (\Gamma_{1}^{R} \check{Q}_R + \Gamma_{1}^{L} \check{Q}_L);\;
{\cal H}_{22}=i\epsilon \tau_z-E_2  - (\boldsymbol{B}_1 \cdot \check{\boldsymbol{\sigma}})\tau_z  
+  \frac{i}{2} (\Gamma_{2}^R \check{Q}_R + \Gamma_{2}^L \check{Q}_L);\\
{\cal H}_{12}&=-\check{\kappa} +\frac{i}{2} \{ \check{\gamma}_{L} \check{Q}_L+\check{\gamma}_R \check{Q}_R \};\;
{\cal H}_{21}=-\check{\kappa}^{\dagger}+\frac{i}{2} \{ \check{\gamma}_{L}^\dagger \check{Q}_L+\check{\gamma}_R^\dagger \check{Q}_R \},
\end{align}

and we turn back to the compact notations

\begin{align}
\check{\kappa},\check{\kappa}^\dagger=\kappa\pm i \boldsymbol{\kappa}\cdot\boldsymbol{\sigma}; \;
\check{\gamma}_{L,R},\check{\gamma}_{L,R}^\dagger=\gamma_{L,R}\pm i \boldsymbol{\gamma_{L,R}}\cdot\boldsymbol{\sigma}
\end{align}
and made use of $Q$ matrices corresponding to two leads
\begin{equation}
\check{Q}_{L,R}=\frac{1}{\sqrt{\epsilon^2 +\Delta^2}}
\begin{bmatrix}
\epsilon& \Delta e^{i\phi_{L,R}}\\
\Delta e^{-i\phi_{L,R}}& -\epsilon
\end{bmatrix}.
\end{equation}


Next goal is to reduce the number of parameters implementing the separation of scales mentioned and implemented for the normal transport.  This is achieved by the following transformation of the determinant
\begin{align}
\ln \det(\check{{\cal H}})=\ln \det(\check{{\cal H}}_{11}-\check{{\cal H}}_{12}\check{{\cal H}}_{22}^{-1}\check{{\cal H}}_{21}) + \ln \det(\check{{\cal H}}_{22})
\end{align}
and implementing $E_2,\Gamma_2 \gg \gamma,\kappa \gg \epsilon, B_2, E_1, \Gamma_1$.

Let us first evaluate $\det(\check{{\cal H}}_{22})$, which is that of a $4\times 4$ matrix with spin structure taken into account. Since we may assume $\epsilon, B_2 \ll \Gamma_2, E_2$ 
the spin structure is trivial and the answer reads 


\begin{align}
\ln \det(\check{{\cal H}}_{22})=2 \ln ( E_2^2 +\frac{1}{4} \Gamma_2^2 )+  2 \ln(1-T_0 \frac{\Delta^2}{\Delta^2 +\epsilon^2 } \sin^2{\phi/2}),
\end{align}
where, as previously, we define $\Gamma_2=\Gamma_{2}^{L}+\Gamma_{2}^{R}$ and $T_0=\Gamma_{2}^{L}\Gamma_{2}^{R}/( E_2^2 +\frac{1}{4} \Gamma_2^2 )$.

The energies of Andreev levels are determined from zeros of this determinant. We recover the well-known expression for the energy of the spin-degenerate Andreev level in a contact with transparency $T_0$,
\begin{equation}
E_{Andr} = \Delta \sqrt{1 - T_0 \sin^2(\phi^/2)}
\end{equation}
The integration of the log of the determinant over the energy gives the expected result for the energy of the ground state,
\begin{equation}
{\cal E} = - E_{Andr}
\end{equation}

Let us turn to evaluation of the rest of the expression. We note that 
\begin{equation}
\check{{\cal H}}^{-1}_{22} = -\frac{E_2  + \frac{i}{2} (\Gamma_{2R} \check{Q}_R + \Gamma_{2L} \check{Q}_L)}{\Big( E_2^2+\frac{\Gamma_2^2}{4}\Big)(1-T_0 s)} 
\end{equation}
where we have introduced a convenient compact notation 
\begin{equation}
s \equiv \frac{\Delta}{\sqrt{\Delta^2 +\epsilon^2}} \sin^2(\phi/2)
\end{equation}
The matrix in the first determinant thus contains a factor $(1-T_0 s)$ in the denominator. Multiplying with this factor cancels $\det(\check{\cal H}_{22})$ so the whole expression can be reduced to the following relatively simple form
\begin{align}
&\ln \det(\check{{\cal H}})=&\label{eq:finalsup}\\
&\ln \det\left( 
(1 -T_0s)(i\epsilon \tau_z-E_1  - (\boldsymbol{B} \cdot \check{\boldsymbol{\sigma}})\tau_z) 
+\Delta E + s \Delta E_S 
+ \frac{i}{2} (\Gamma^R(s) \check{Q}_R + \Gamma^{L}(s) \check{Q}_L) +\frac{i}{4} \boldsymbol{\Gamma}\cdot \check{\boldsymbol{\sigma}} (\check{Q}_L \check{Q}_R - \check{Q}_R \check{Q}_L)\right)&,\nonumber
\end{align}
where
\begin{align}
\Gamma^{L,R}(s) = \Gamma^{L,R} + s \Gamma_S^{L,R}.
\end{align}
The parameters $\Gamma_{L,R}$, $\Delta E$, $\boldsymbol{\Gamma}$ have been already defined in our consideration of normal transport. The compact description of superconducting transport brings three additional parameters
\begin{align}
\Delta E_S = \frac{ - E_2 C_7 + \kappa C_8 + \boldsymbol{\kappa} \cdot \boldsymbol{C}_9}{E_2^2 + \Gamma_2^2/4} ;\;
\Gamma_S^{L}= - T_0 \Gamma_1^L + \frac{\Gamma_R (\gamma_L^2 + \boldsymbol{\gamma}_L^2)}{E_2^2 + \Gamma_2^2/4} ;\;
\Gamma_S^{R}= - T_0 \Gamma_1^R + \frac{\Gamma_L (\gamma_R^2 + \boldsymbol{\gamma}_R^2)}{E_2^2 + \Gamma_2^2/4}.
\end{align}
Here, $\Delta E_S$ is a part of the expression (\ref{eq:RX}) for $RX$ but is an independent parameter. 

Since both normal and superconducting transport originate from the same scattering matrix, there are many examples when the parameters characterizing the superconducting transport can be directly determined from the results of normal transport measurements, a single channel with transparency $T_0$ being the simplest one. The presence of the additional parameters $\Delta E_S$, $\Gamma_S^{L,R}$is therefore rather disappointing: we cannot predict superconducting transport exclusively from the results of normal transport measurements and have to rely on model assumptions.

Let us outline the physical meaning of the overall structure of the expression (\ref{eq:finalsup}). The first term is a product of the terms whose zeros give the Andreev level in the transport channel and energy level in an isolated localized state, the product indicate that these levels are independent. The rest of the terms thus describe the hybridization of these levels. Note that the terms with $\Delta E$ cannot be cancelled by a shift of $E_1$, so in distinction from the normal case, are active in the presence of superconductivity. The terms with $\Gamma(s)$ are similar to tunnel decay terms in Eq. \ref{eq:H11}, in distinction from normal case the presence of the second dot does not just renormalize $\Gamma$. The last term describes spin-orbit effect and is proportional to the same vector $\boldsymbol{\Gamma}$ as in the normal case. In distinction from all other terms, it is odd in the phase difference since it is proportional to the commutator of two $\check{Q}$. The combination of this term and that with magnetic field results in a shift of the minimum of the phase-dependent energy from $0$ or $\pi$ positions.


\subsection{Numerical details}
\label{sec:numerics}
In this subsection, we provide the overall strategy and details of our numerical calculations.

We postpone the discussion of self-consistency assuming that we already know $E_1$ and $\boldsymbol{B}$. To find the phase-dependent energy, we have to integrate the log of the determinant over $\epsilon$. We compute directly the determinant of $8 \times 8$ matrices implementing the difference of scales numerically.  
For quick computation at each energy, we represent the matrix $\check{{\cal H}}$ as a sum over various scalar functions of $\epsilon$,

\begin{equation}
\check{{\cal H}} =\check{A}+\epsilon \check{B} +\frac{\epsilon}{\sqrt{\epsilon^2+\Delta^2}} \check{C} + \frac{\Delta}{\sqrt{\epsilon^2+\Delta^2}} \check{D}(\phi_L,\phi_R)
\end{equation}
where the matrices $\check{A} - \check{D}$ do not depend on $\epsilon$ and only $\check{D}$ depends on the superconducting phase.
 We define the function of $\epsilon$ as 
 $\log(\det(\check{{\cal H}}(\epsilon,\phi=0))-\log(\det(\check{{\cal H}}(\epsilon,\phi=0))$
 and integrate using scipy.quad. Direct evaluation of the sum over discrete equidistant $\epsilon$ the interval of the order of $5 \Delta $ provides comparable numerical efficiency and accuracy. 

As mentioned, we treat interaction self-consistently, as the interaction-induced shift in $E_1$ and $\boldsymbol{B}$. The self-consistency equations read as

\begin{align}
\tilde{E_1}=E_1+U n(\tilde{E_1},\tilde{\boldsymbol{B}}); \;
\tilde{\boldsymbol{B}}=\boldsymbol{B}-U \boldsymbol{n}(\tilde{E_1},\tilde{\boldsymbol{B}})
\end{align}

,where the average number of particles on the dot are given by derivatives of the total energy
$N=(\partial_{E_1} {\cal E} +1) $ and $\boldsymbol{n}=\partial_{\boldsymbol{B_1}}{\cal E}$.
We compute these derivatives integrating the analytical derivatives of $\det(\check{{\cal H}}(\epsilon)$ over $\epsilon$. These integrals may converge at $\epsilon \gg \Delta$ provided $E_{1},\boldsymbol{B} \gg \Delta$. An adaptive grid of discrete $\epsilon$ could be chosen to speed up the evaluation, yet using scipy.quad suffices for our purposes.

To solve the self-consistency equations, we implement a root-finding  minimization algorithm minimizing the function  $F=f^2+|\boldsymbol{f}|^2$, where $f, \boldsymbol{f}$ are defined as 
\begin{align}
f=\tilde{E_1}-E_1-U n(\tilde{E_1},\tilde{\boldsymbol{B}});\;
\boldsymbol{f}=\boldsymbol{\tilde{B}}-\boldsymbol{B}+U \boldsymbol{n}(\tilde{E_1},\tilde{\boldsymbol{B}}),
\end{align}
and checking if the minimum is achieved at $F=0$. 
Alternatively, we can make use of the fact that the solutions of the self-consistency equations correspond to the extrema of the following energy functional 
\begin{equation}
E_{T}(\tilde{E}_1,\tilde{\boldsymbol{B}})= {\cal E}(\tilde{E}_1,\tilde{\boldsymbol{B}}) -\frac{(E_1 - \tilde{E}_1)^2}{2 U} + \frac{(\boldsymbol{B}- \tilde{\boldsymbol{B}})^2}{2 U}
\end{equation}
Unfortunately, this energy function is not bounded, and the extrema required are rather saddle points than minima. However, they can be found, for instance, by maximization of the function in $\tilde{E}_1$ and subsequent minimization in $\tilde{\boldsymbol{B}}$.


The Andreev bound states are given by the zeros of the  determinant at imaginary $\epsilon$ (thus real energy $E = i\epsilon$)
in the interval $(0,\Delta)$. We find these roots by minimizing $\det(\check{{\cal H}}(\epsilon))^2$ and checking if the minimum is achieved at zero. but because of the different scales in the problem and existence of a big scale, we first try to find an equivalent matrix, the determinant of which is more efficiently minimized numerically. 
Typically, there are multiple Andreev states, so we subdivide the interval $(0,\Delta)$ to find them all.

\subsection{Superconducting transport examples}
\label{sec:supexamples}

In this subsection, we present the examples of numerical evaluation of supercurrent and Andreev state energies in our model, for 3 sets of parameters.
The $\pi$ state is achieved if $E_T(\phi=\pi) - E_T(\phi=0)\equiv E_\pi <0$. Contrary to our initial expectations, it is rather difficult to achieve such inversion for an arbitrary parameter set at high transmission $T_0$. It is rather easy to find $0-\pi$ transitions at low $T_0$. The following examples provide interesting illustrations of rich physics captured by the model.

In this subsection and in all plots, we measure the energies, decay rates and the current $I/2e$ in units of $\Delta$. 

\begin{figure*}[hb]
	\centering
	\begin{subfigure}{0.45\textwidth}
		\includegraphics[width=\textwidth]
{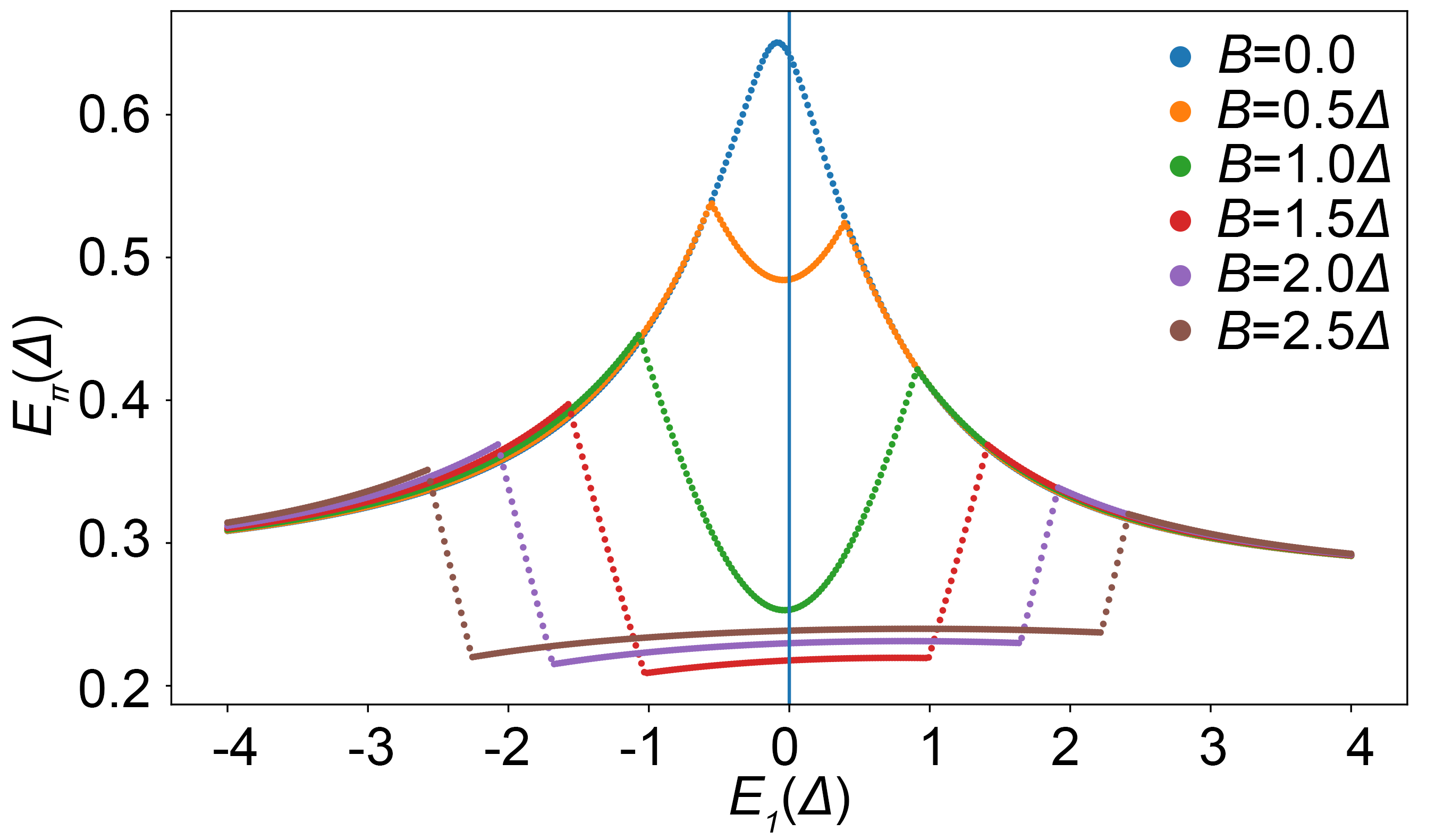}
			\caption{$E_\pi \equiv E_T(\phi=\pi) - E_T(\phi=0)$ versus $E_1$ at several values of magnetic field.}
			\label{fig:E_T_symmetric}
		\end{subfigure}
		\begin{subfigure}{0.45\textwidth}
			\includegraphics[width=\textwidth]
{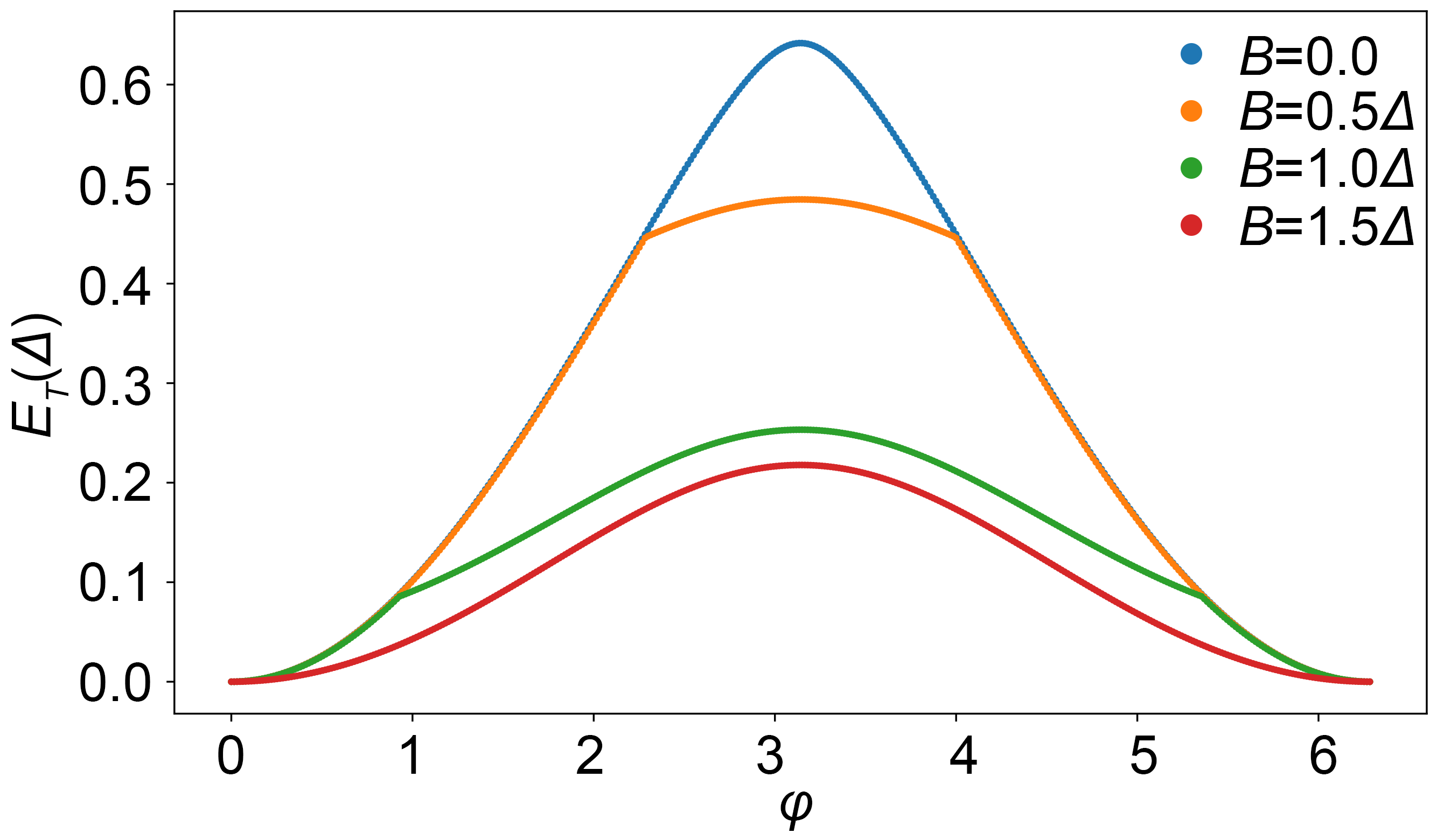}
			\caption{The phase-dependent part of energy $E_T \equiv E_T(\phi) - E_T(\phi=0)$ at $E_1=0$ and several values of magnetic field.}
			\label{fig:E_T_symmetric1}
		\end{subfigure}
		\begin{subfigure}{0.45\textwidth}
			\includegraphics[width=\textwidth ]
{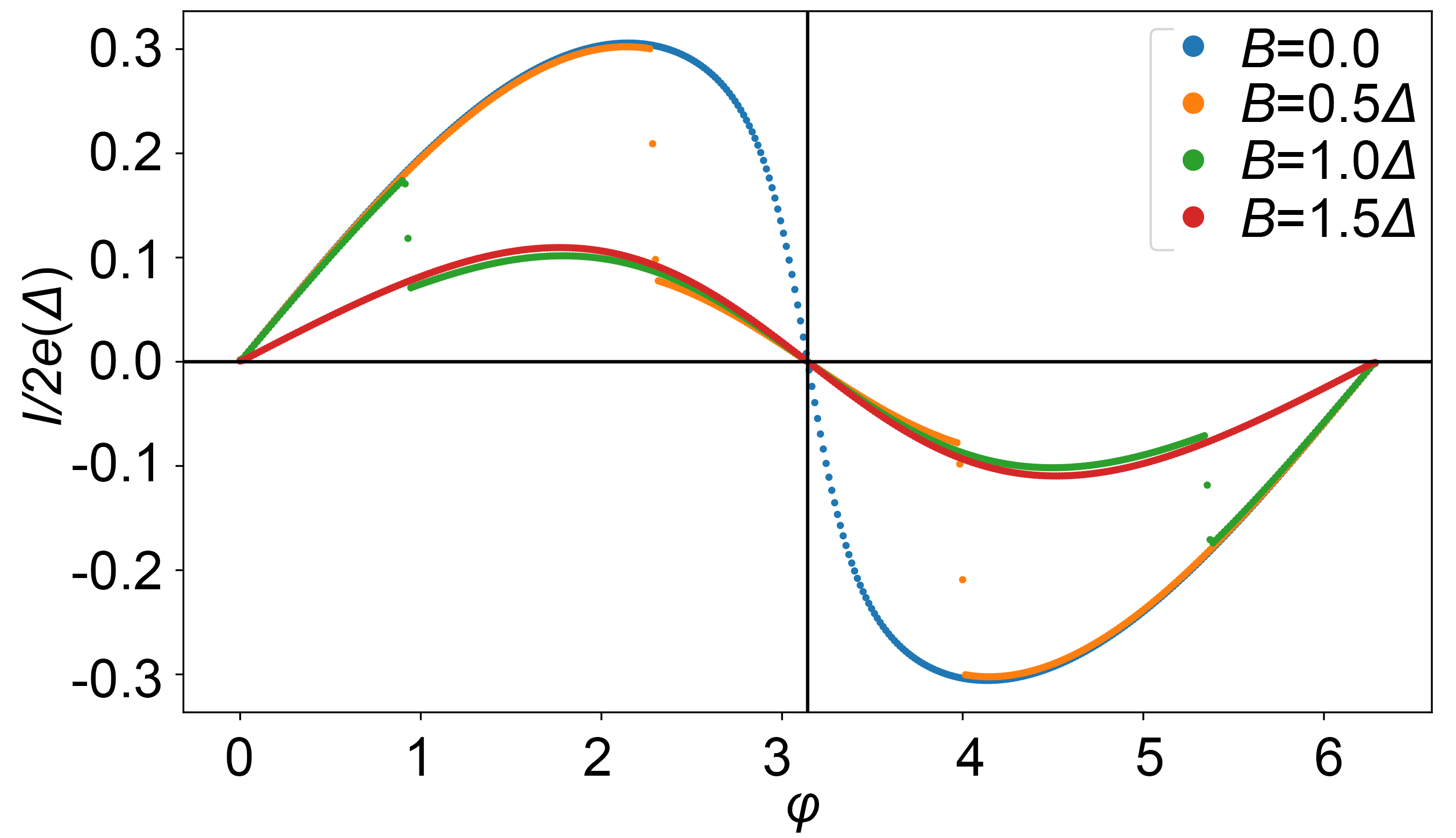}
			\caption{The phase dependence of the superconducting current at $E_1=0$ for several values of $B$.}
			\label{fig:I_symmetric}
		\end{subfigure}
		\begin{subfigure}{0.45\textwidth}
			\includegraphics[width=\textwidth]
{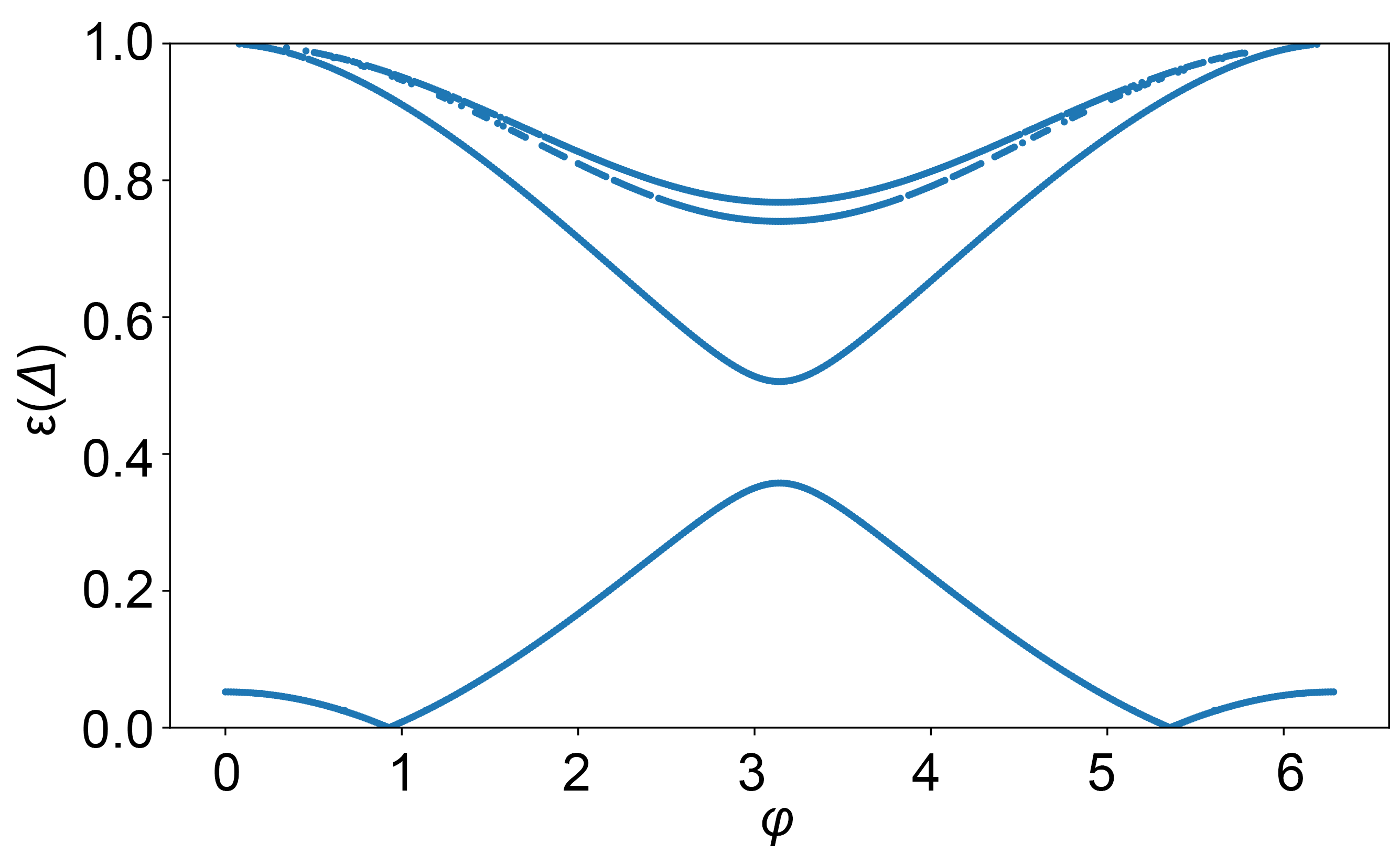}
			\caption{The phase dependence of ABS energies at $E_1=0$ and $|B|=1$.}
			\label{fig:ABS_symmetric}
		\end{subfigure}
	\begin{subfigure}{0.45\textwidth}
		\includegraphics[width=\textwidth]
{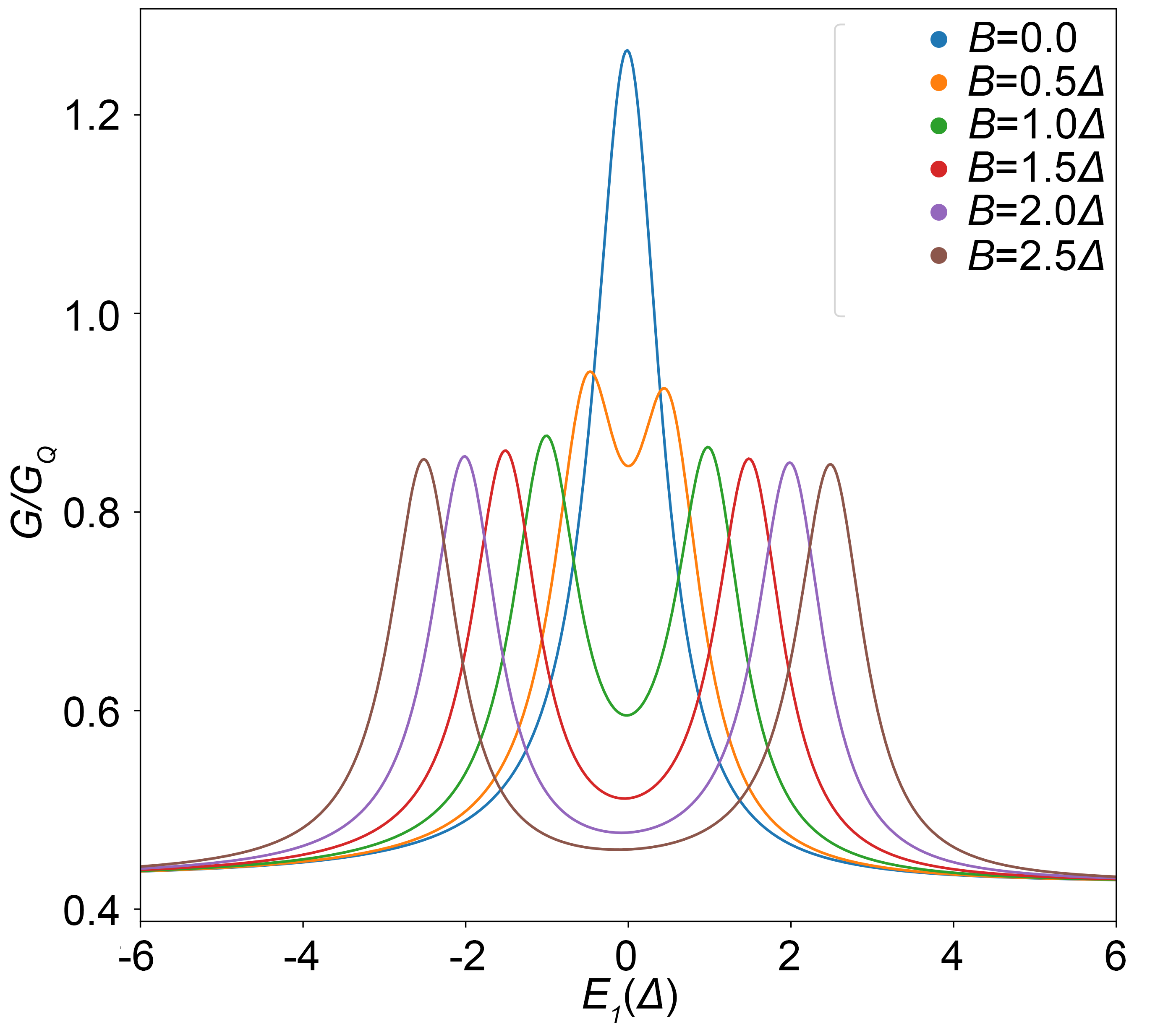}
		\caption{Normal zero-voltage conductance versus $E_1$ at several values of magnetic field.}
		\label{fig:ABS_symmetric2}
	\end{subfigure}
	\caption{Example A. Resonant transmission, moderate channel transmission. No SO coupling. }
	\label{fig:Hristo0}
	\end{figure*}

{\it Example A} (Fig. \ref{fig:Hristo0})
Here, we disregard spin-orbit coupling and interaction. The parameters are $\Gamma_2^L,\Gamma_2^R,E_2 = A(1.9,1.9,-2)$, $\kappa, \gamma_L,\gamma_R = \sqrt{A}(0.4,0.1,0.1)$, $\Gamma_1^L, \Gamma_1^R = 1.2, 0.9$
and correspond to $T_0 = 0.47$, $\Gamma_L =1.3 , \Gamma_R =0.96, \Gamma =2.26$.
As we can see from the normal conductance traces presented in Fig. \ref{fig:ABS_symmetric2}, for this parameter set we have the resonant transmission accompanied by very weak Fano asymmetry. The resonant transmission peak splits upon increasing the magnetic field. 

Actually, this set illustrates an unsuccessful attempt to achieve a pair of $0-\pi$ transitions. This is seen from the plots in Fig. \ref{fig:E_T_symmetric} that give the (gate-voltage) traces of $E_\pi$ at various magnetic fields. Zero-field trace peaks near the centre of conductance peak indicating the enhancement of supercurrent upon increasing the transmission, and saturates at finite value at $|E_1| \gg 1$: this saturation is achieved for all magnetic fields. Upon increasing the magnetic field the value of $E_T$ near the resonance gets down. It becomes smaller than the saturated value at $B > 0.8$. It seems it has a chance to pass zero manifesting $0-\pi$ transitions. However, this does not happen: the tendency changes and the minimum of $E_T$ starts to increase at $B > 1.5$. Prominent features in the plots are sharp cusps in energy dependence. They indicate the crossings of Andreev states with zero energy that, for the features in this plot, are located at $\phi=\pi$ and corresponding gate voltages.

Let us set $E_1=0$ and look at the phase dependence of energy (Fig. \ref{fig:E_T_symmetric1}) for a set of magnetic fields. Here we also see the cusps corresponding to the crossings at certain values of the phase. Superconducting currents plotted in Fig. \ref{fig:I_symmetric} are obtained by numerical differentiation of the energy, so the cusps become jumps, the discontinuities of the current. The zero-field curve is prominently non-sinusoidal as expected for high transmission at this value of $E_1$. The current becomes smaller tending to almost sinusoidal curve at high magnetic fields upon increasing magnetic field, but does not get inverted. At intermediate fields, the current jumps between non-sinusoidal and sinusoidal curve.

In Fig. \ref{fig:ABS_symmetric} we show the phase dependence of ABS energies for $E_1=0$ and $|B_1|=1$. We see four spin-split ABS counting from down up. Eventually, the picture of ABS demonstrates little interference between the transport channel and the localized state. The third and the fourth curves are close to $E_{Andr}$ for $T_0=0.47$ and are thus associated with the transport channel, their spin-splitting $\simeq 0.1$ is small coming from the interference. The first and the second state are associated with the dot. The spin splitting is thus big: the first curve  looks like the second curve shifted by $\simeq 1$ downwards, with the part shifted to negative energy being flipped to positive ones. This also explains sharp cusps in the first curve.

Although in our model the phase-dependent energy is not a minus half-sum of ABS energies as it would be for energy-independent transmission, we can use this sum for qualitative estimations. With this, the half-sum of the first and second energies would result in an inverted supercurrent, but the half-sum of the third and fourth states, that is, the contribution of the transport channel, adds to the balance a usual supercurrent of slightly bigger amplitude.

\begin{figure*}
	\centering
	\begin{subfigure}{0.45\textwidth}
		\includegraphics[width=\textwidth]
{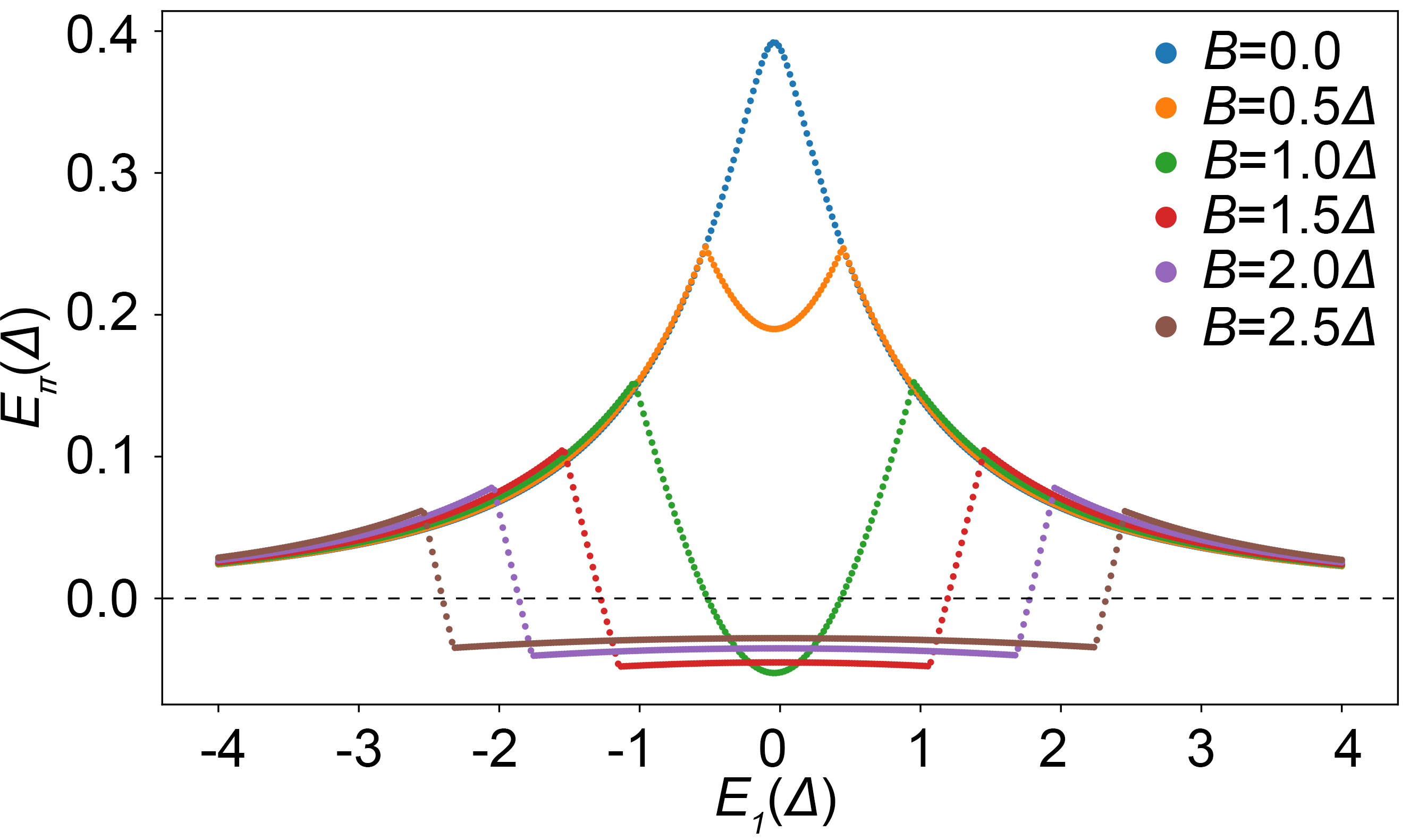}
		\caption{$E_\pi$ versus $E_1$ at several values of magnetic field.}
		\label{fig:E_T_inversion}
	\end{subfigure}
	\begin{subfigure}{0.45\textwidth}
		\includegraphics[width=\textwidth ]
	{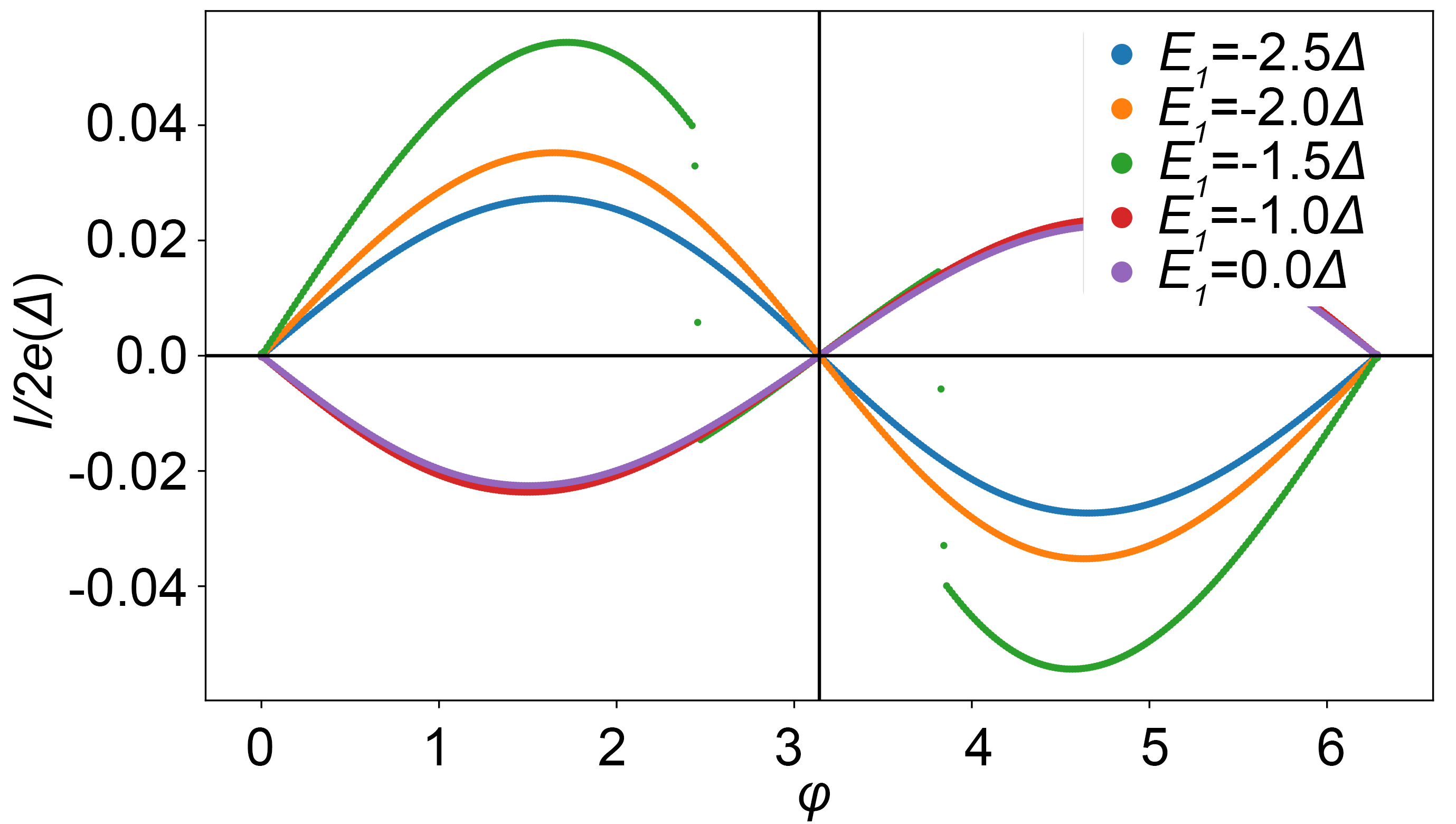}	
		\caption{The phase dependence of the superconducting current at $|B|=1.5$ for several values of $E_1$.}
		\label{fig:I_inversion}
	\end{subfigure}
	\begin{subfigure}{0.45\textwidth}
		\includegraphics[width=\textwidth]
		{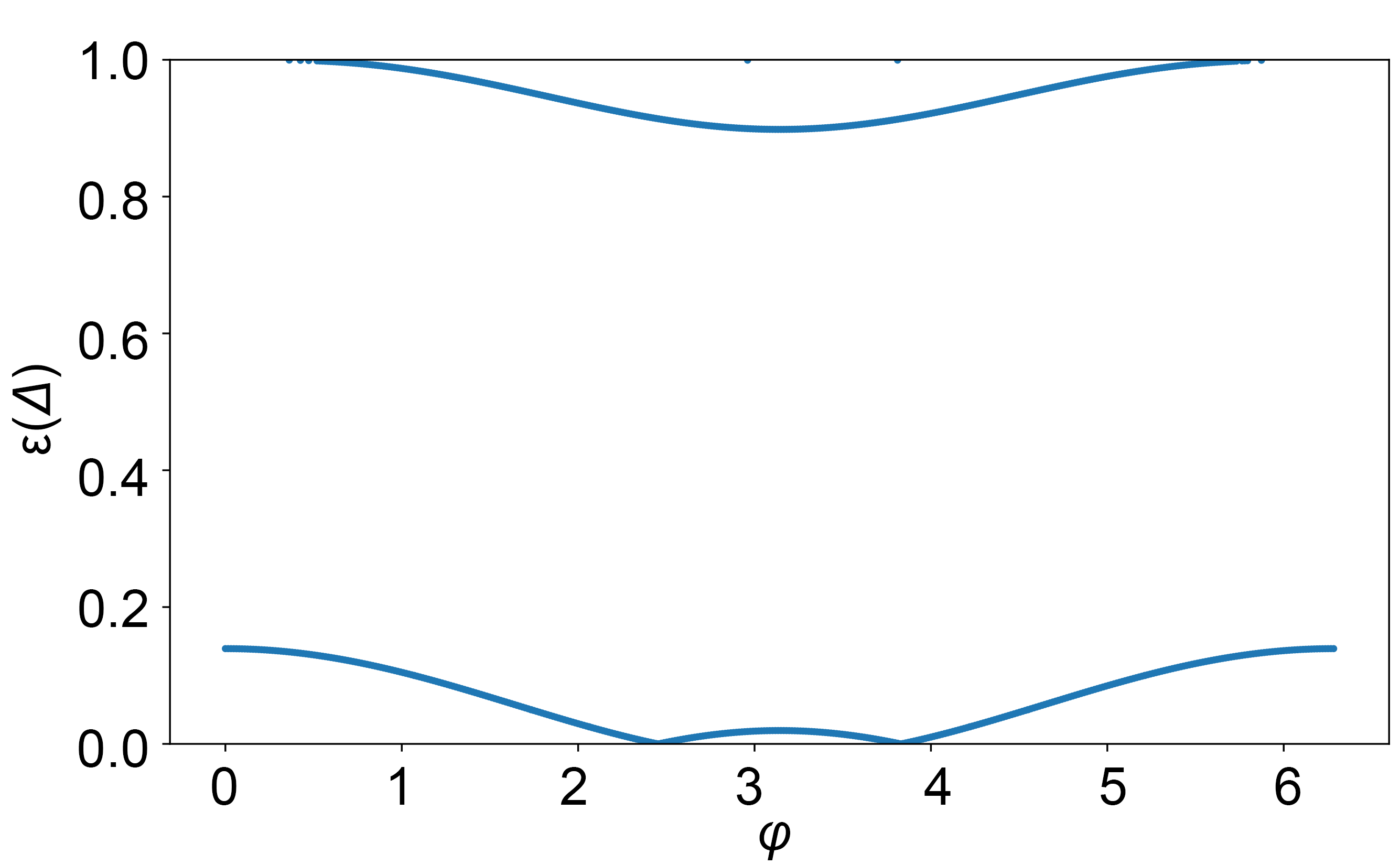}
		\caption{The phase dependence of ABS energies at $E_1=-1.5$ and $|B|=1.5$.}
		\label{fig:ABS_inversion}
	\end{subfigure}
	\begin{subfigure}{0.45\textwidth}
	\includegraphics[width=\textwidth]
	{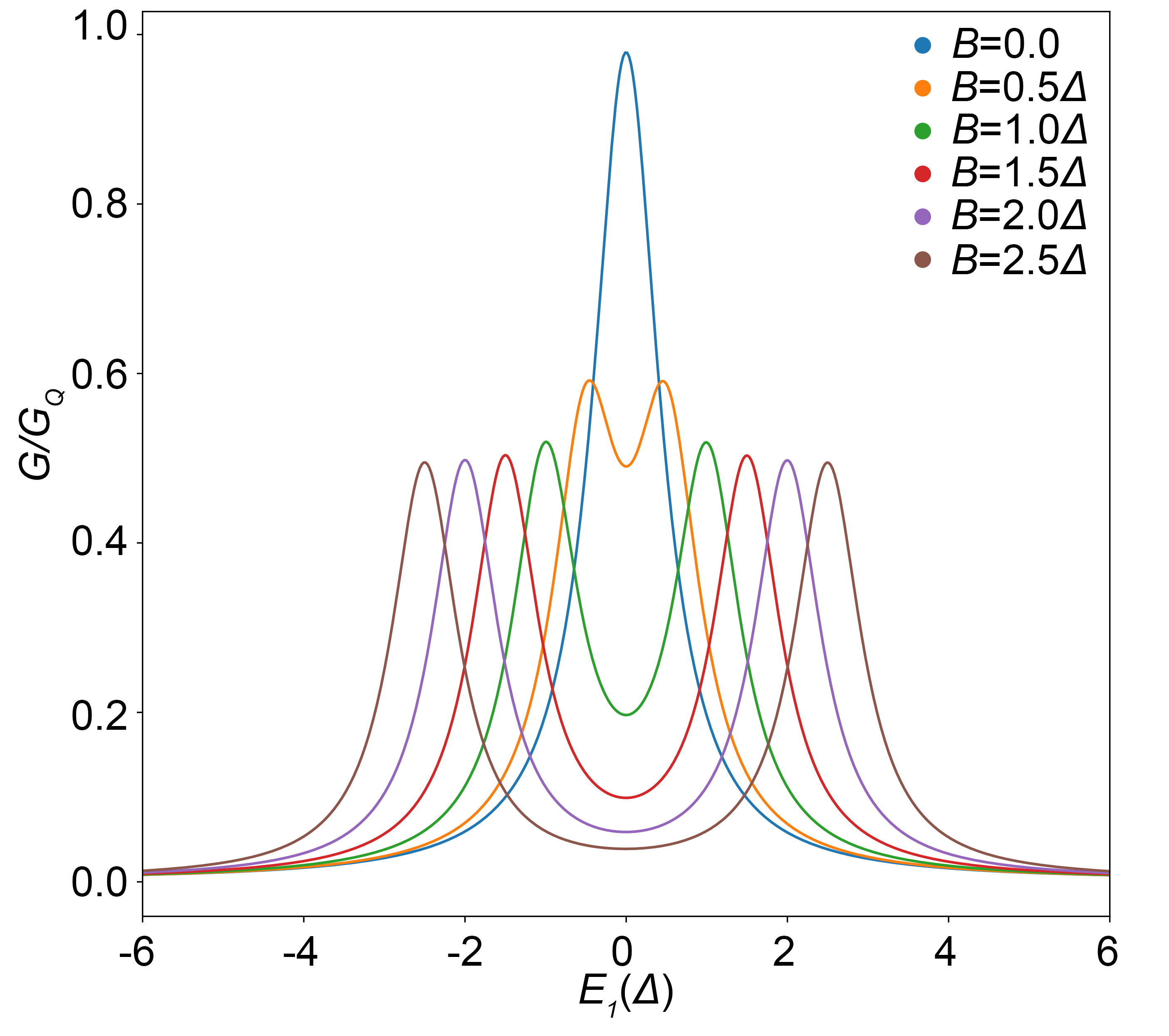}
	\caption{Normal zero-voltage conductance versus $E_1$ at several values of magnetic field.}
	\label{fig:Normal_conductance_symmetry}
\end{subfigure}
\caption{Example B. Resonant transmission, low channel transmission. No SO coupling. A pair of $0-\pi$ transitions occurs at $|B|>0$.}
\label{fig:Hristo1}
\end{figure*}

{\it Example B}. (Fig. \ref{fig:Hristo1}) This inspired us to check if the $0-\pi$ transitions can be achieved at very low transmission of the transport channel. We have taken the following set of parameters $\Gamma_2^L,\Gamma_2^R,E_2 = A(0.2,1.5,-15)$, $\kappa, \gamma_L,\gamma_R = \sqrt{A}(0.8,0.1,0.1)$, $\Gamma_1^L, \Gamma_1^R = 1.1, 0.91$. For this choice, $T_0\simeq 10^{-3}$  $\Gamma_L=1.1, \Gamma_R = 0.91, \Gamma=2.01$. 
The normal conductance traces (Fig. \ref{fig:Normal_conductance_symmetry}) 
show a classical scenario of resonant transmission that saturates to almost zero far from the resonance. 
 
The check was successful.
We plot the traces of  $ E_T \equiv E_T(\phi=\pi) - E_T(\phi=0)$ for various magnetic fields in Fig. \ref{fig:E_T_inversion}. The traces look like those in Fig. \ref{fig:E_T_symmetric} except the shift downwards by $\simeq 0.25$. Owing to this, $E_T$ is negative for $B> 0.8$ in an interval of gate voltages that increases with $B$, $0-\pi$ transitions are at the ends of the interval. 

We plot the phase dependence of the supercurrent for $|B|=2$ and various $E_1$ in Fig. \ref{fig:I_inversion}. The $0-\pi$ transitions at this field take place at $E_1 \approx \pm 1.25$. In accordance with this, the almost sinusoidal curves at $E_1 = -2.5, 2$ are of positive amplitude while those at $E_1 = 0, -1$ are of negative one. Note a rather low value $\simeq 0.02$ of the maximum "negative" current, almost 25 times smaller than the maximum value of the current in a single transport channel. An interesting curve is found close to the transition, at $E_1 = -1.5$. Here, the current jumps between sin-like curves of positive and negative amplitude. The total integral of the current between $0$ and $\pi$ is still positive, so $E_\pi >0$. 

An example of the phase dependence of ABS energies is given in Fig. \ref{fig:ABS_inversion}. Since the transmission of the channel is very low, we see only two spin-split ABS. The upper one is close to the gap edge, and eventually merges with continuous spectrum at $\phi \approx 0.6, 2\pi - 0.6$. The lower one is close to zero, and exhibits two zero crossings at $\phi \approx \pi \pm 0.65$ corresponding to the discontinuities in corresponding curve in Fig. \ref{fig:I_inversion}.

The example presented concerns practically zero background transmission, which is not experimental situation. The Fig. 4 in the main text presents the results at small but finite transmission $T_0 \approx 0.3$. 

\begin{figure*}
	\centering
	\begin{subfigure}{0.45\textwidth}
		\includegraphics[width=\textwidth]
		{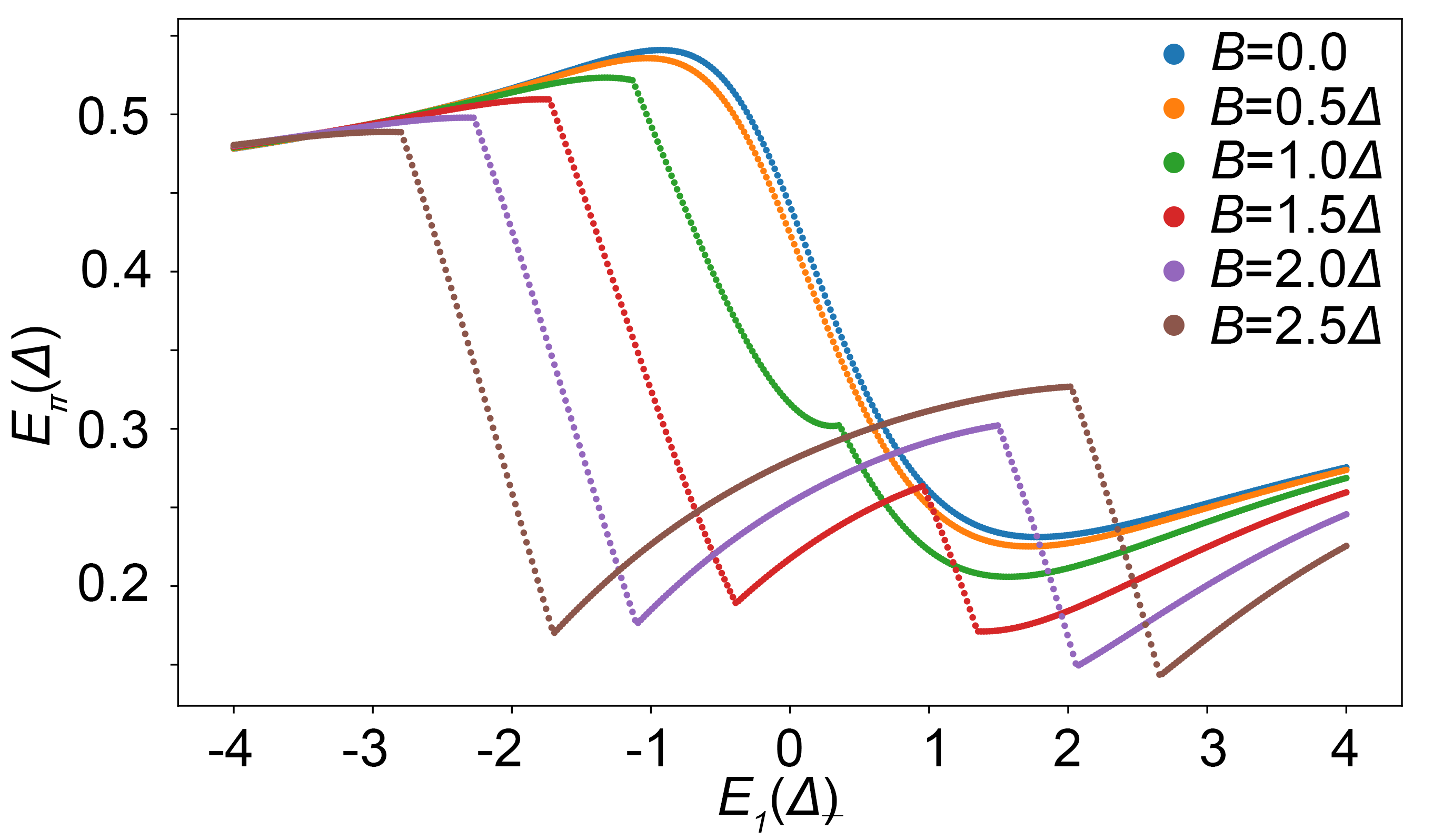}
		\caption{$E_\pi$ versus $E_1$ at several values of magnetic field.}
		\label{fig:E_T_SO_perp}
	\end{subfigure}
	\begin{subfigure}{0.45\textwidth}
		\includegraphics[width=\textwidth]
		{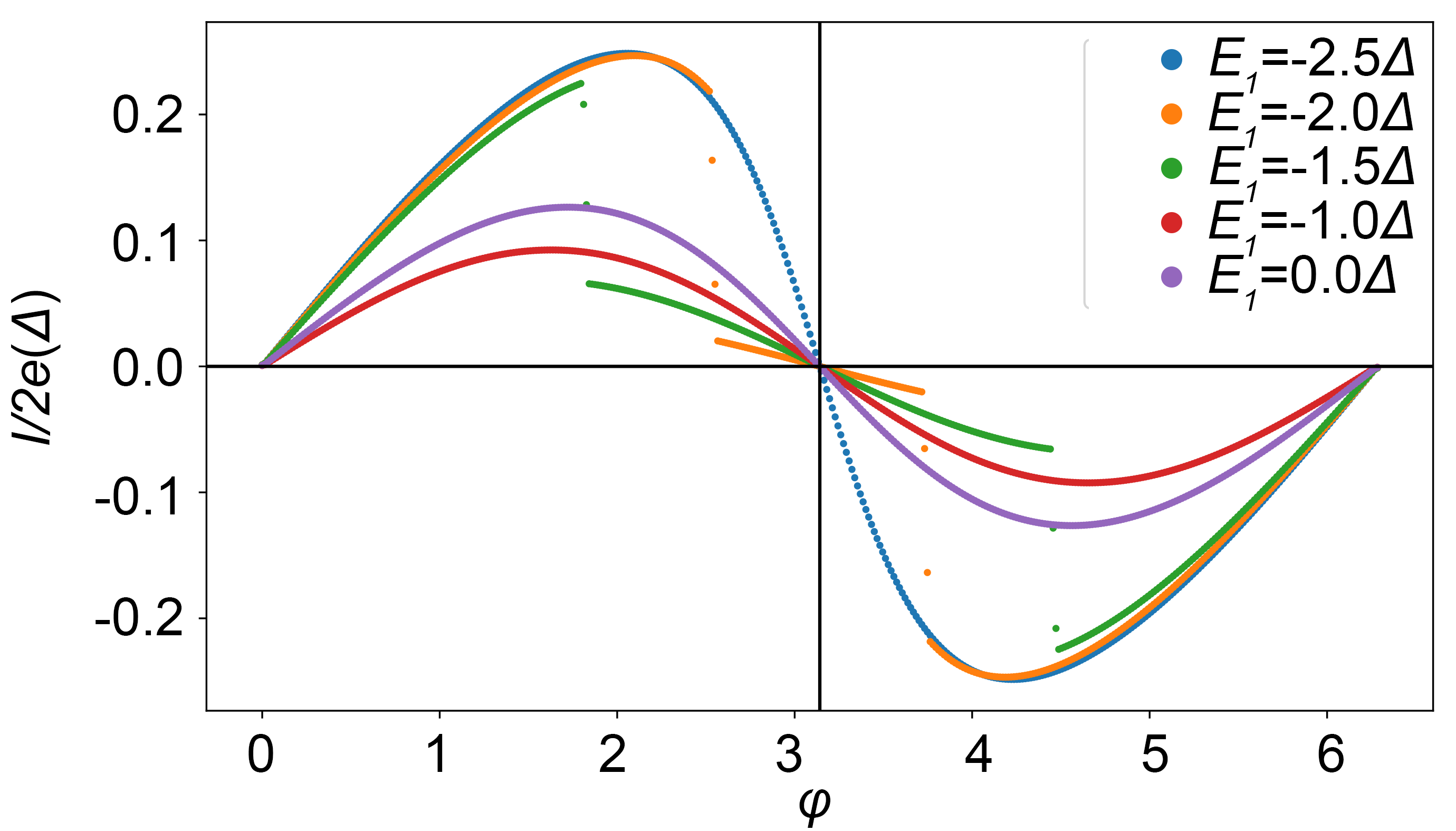}
		\caption{The phase dependence of the superconducting current at $B=2$ for several values of $E_1$.}
		\label{fig:I_SO_perp}
	\end{subfigure}
	\begin{subfigure}{0.45\textwidth}
		\includegraphics[width=\textwidth ]
		{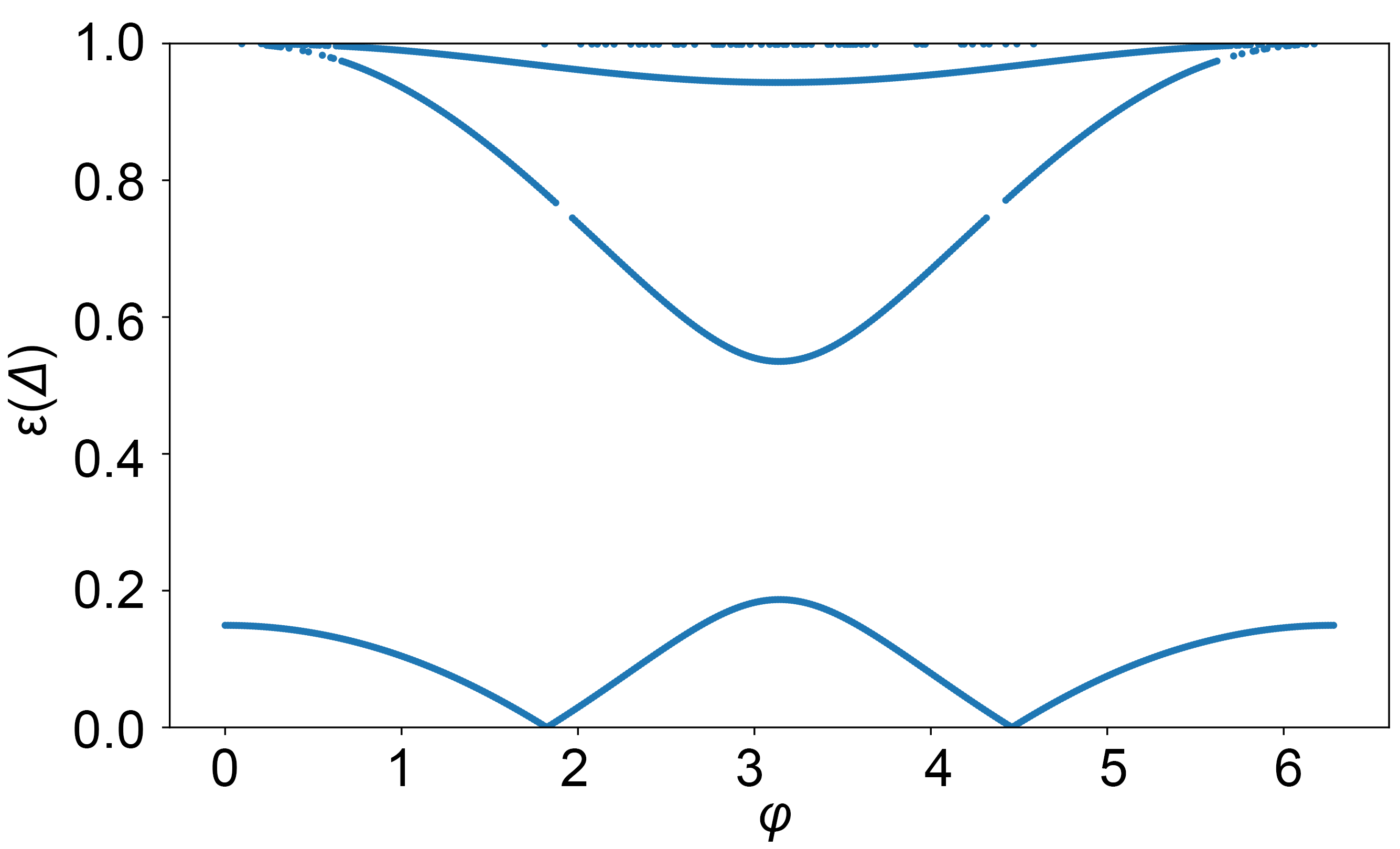}
		\caption{The phase dependence of ABS energies at $E_1=-1.5$ and $B=2$.}
		\label{fig:ABS_SO_perp}
	\end{subfigure}
	\begin{subfigure}{0.45\textwidth}
	\includegraphics[width=\textwidth ]
	{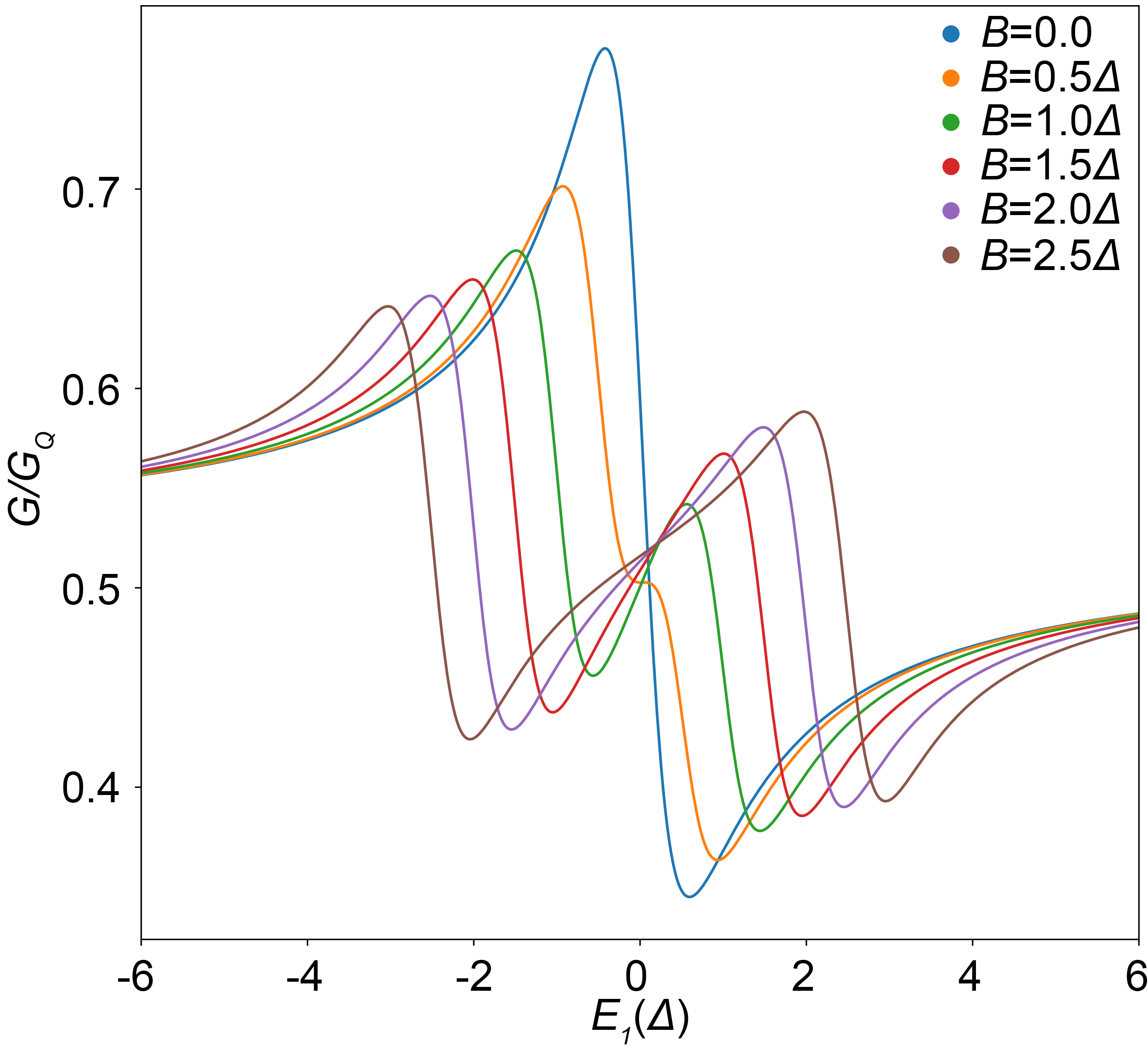}
	\caption{Normal zero-voltage conductance versus $E_1$ at several values of magnetic field.}
	\label{fig:normal_conductane_Perp_SO}
\end{subfigure}
\caption{Example C. Well-developed Fano features, moderate SO coupling. Magnetic field $ \boldsymbol{B}\perp \boldsymbol{\Gamma}$}
\label{fig:Hristo2}
\end{figure*}

\begin{figure*}
	\centering
	\begin{subfigure}{0.45\textwidth}
		\includegraphics[width=\textwidth]
		{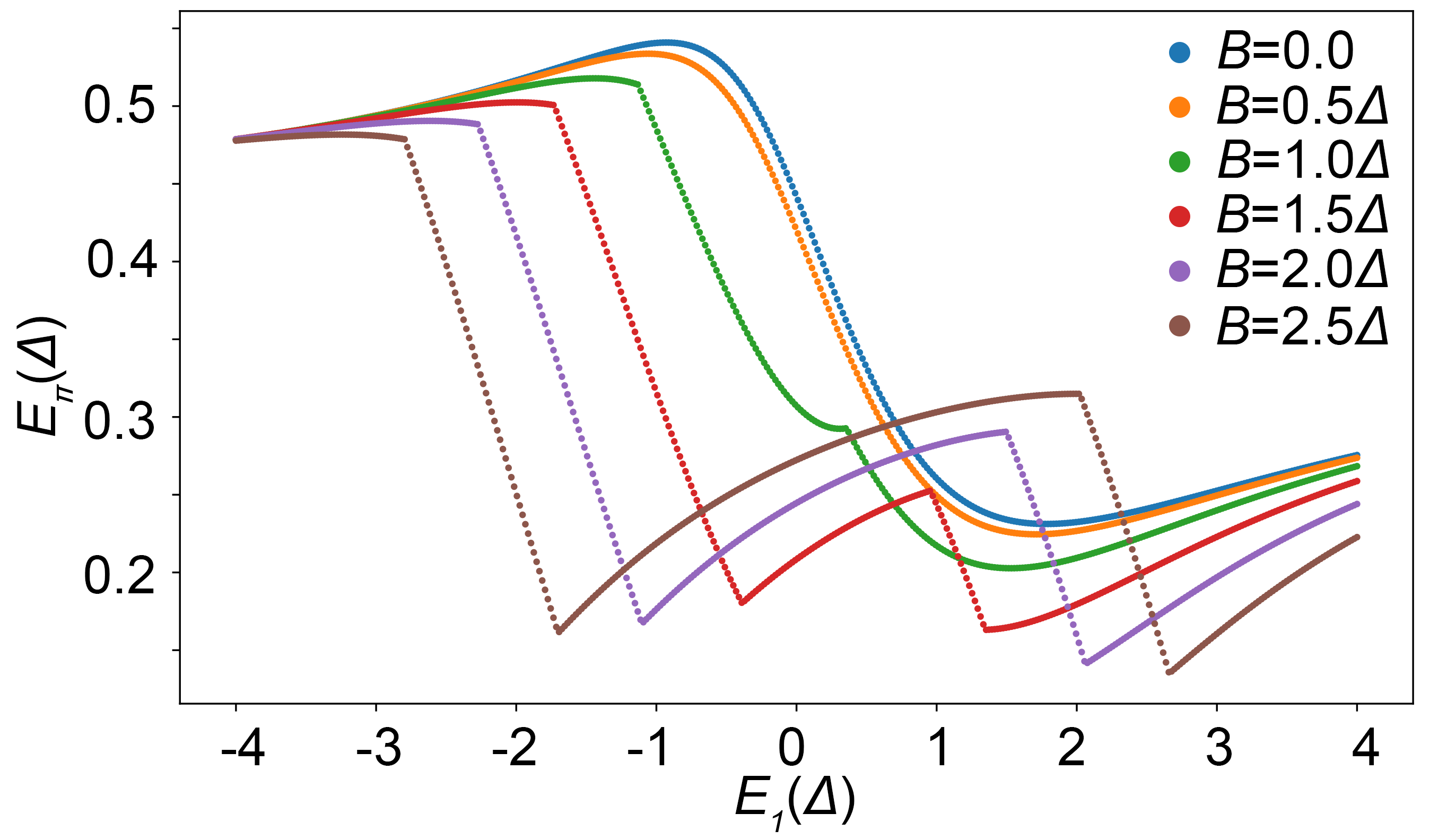}
		\caption{$E_\pi$ versus $E_1$ at several values of magnetic field.}
		\label{fig:E_T_SO_along}
	\end{subfigure}
	\begin{subfigure}{0.45\textwidth}
		\includegraphics[width=\textwidth]
		{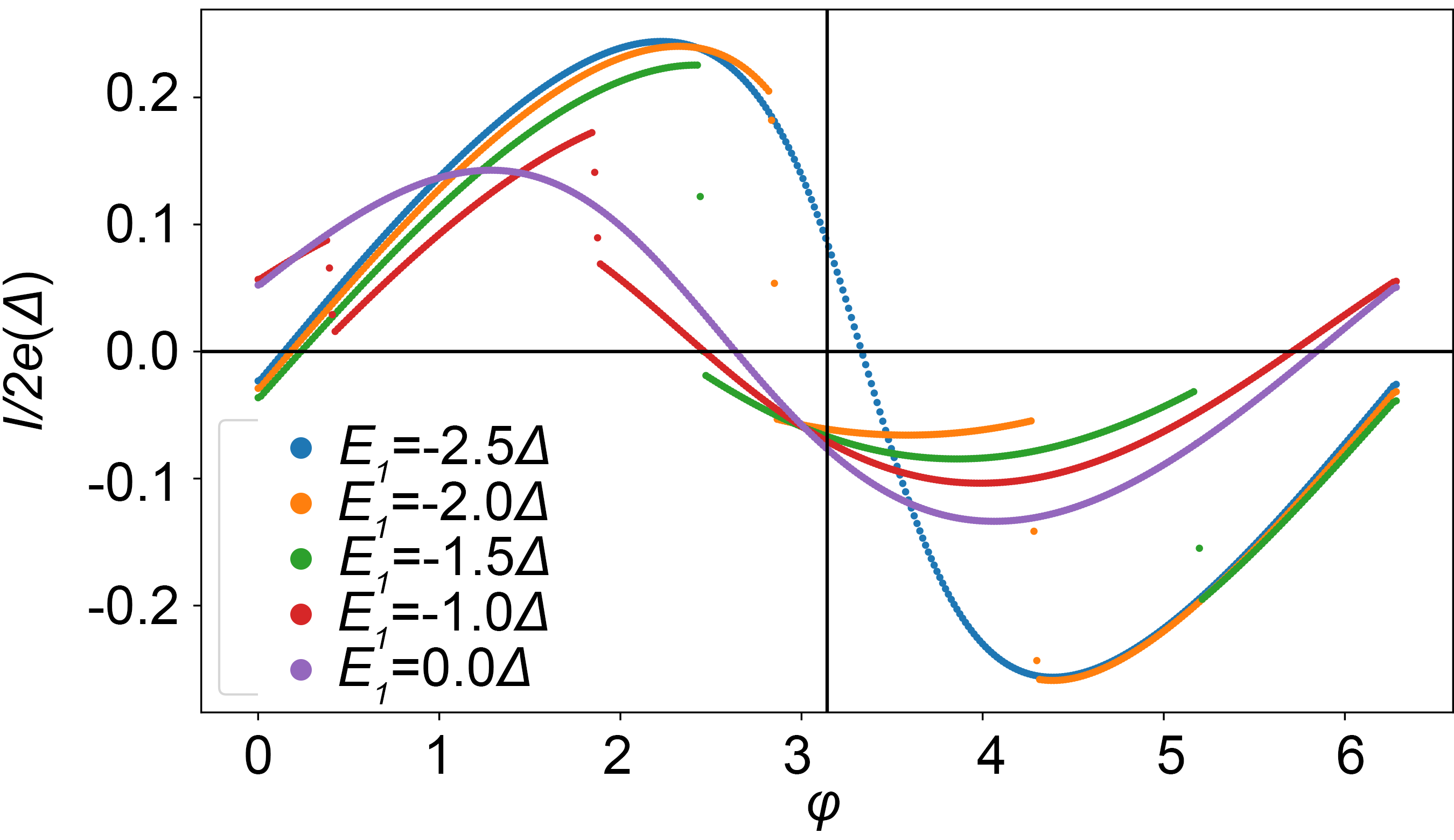}
		\caption{The phase dependence of the superconducting current at $B=1.5$ for several values of $E_1$.}
		\label{fig:I_SO_along}
	\end{subfigure}
	\begin{subfigure}{0.45\textwidth}
		\includegraphics[width=\textwidth ]
		{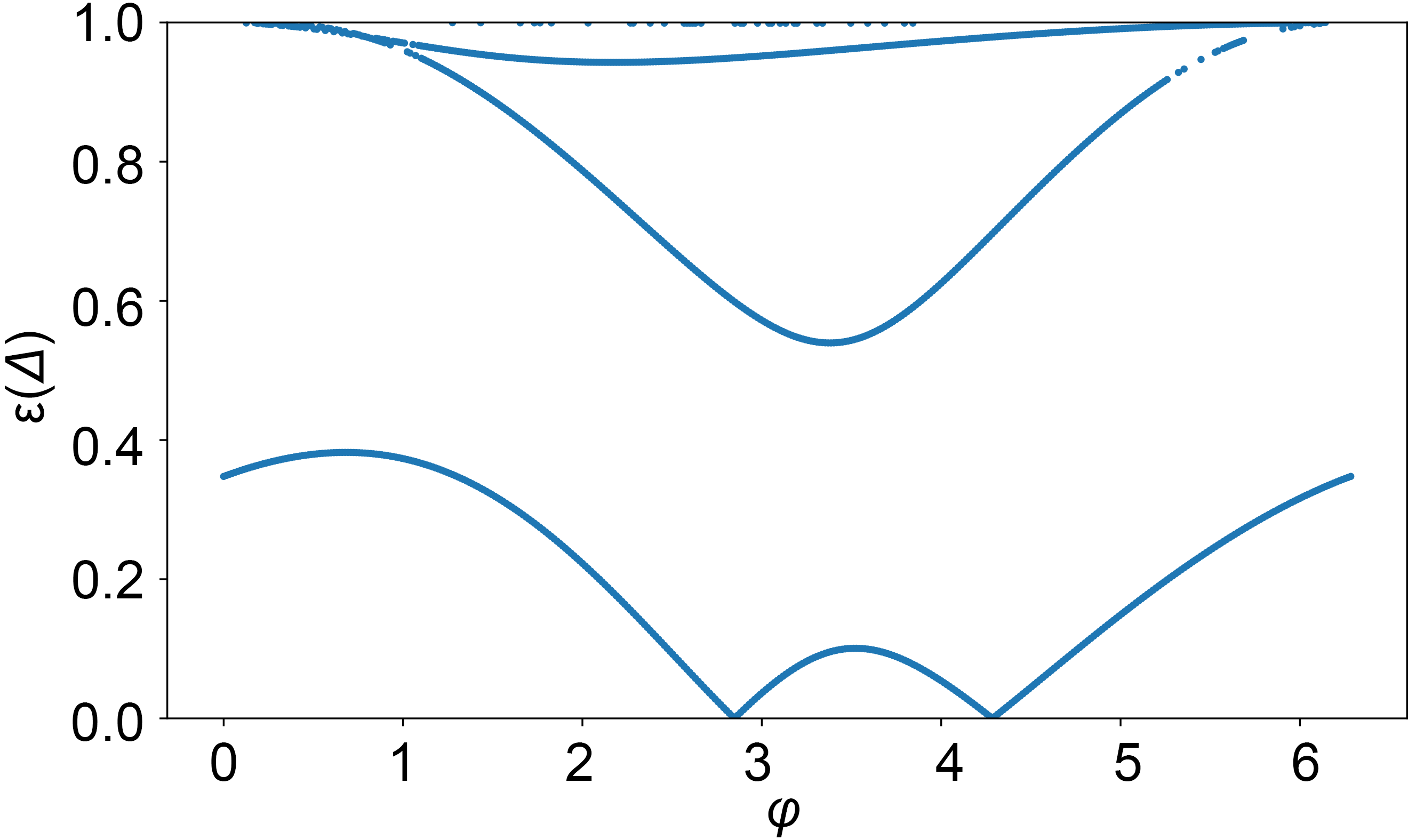}
		\caption{The phase dependence of ABS energies at $E_1=-2$ and $|B|=2$.}
		\label{fig:ABS_SO_along}
	\end{subfigure}
	\begin{subfigure}{0.45\textwidth}
	\includegraphics[width=\textwidth ]
	{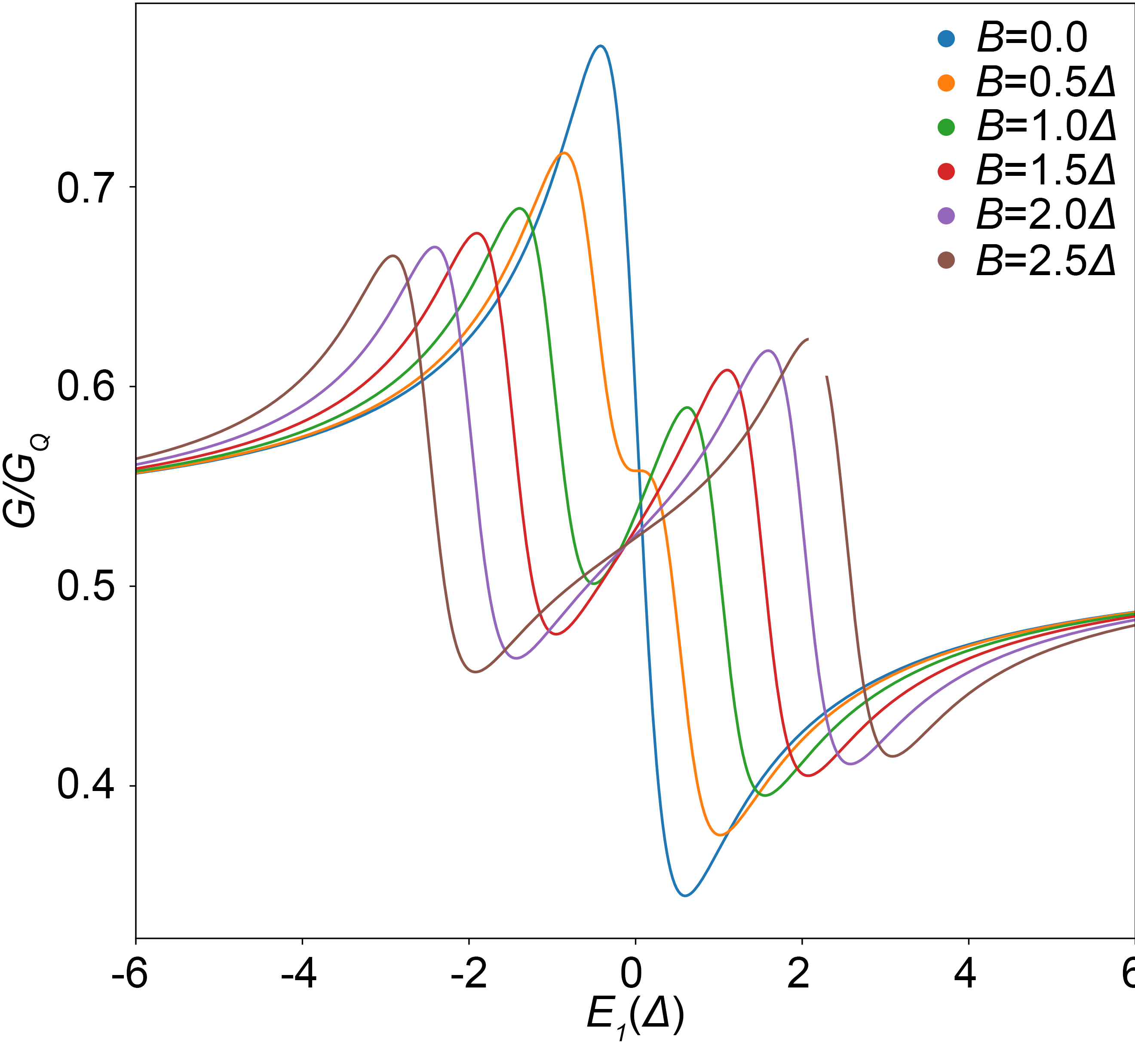}
	\caption{Normal zero-voltage conductance versus $E_1$ at several values of magnetic field.}
	\label{fig:Normal_Conductance_along}
\end{subfigure}
\caption{Example C. Well-developed Fano features, moderate SO coupling. No interaction. Magnetic field $ \boldsymbol{B} \parallel \boldsymbol{\Gamma}$. Pronounced asymmetry in $\phi$.}
\label{fig:Hristo3}
\end{figure*}

{\it Example C.} (Figs. \ref{fig:Hristo2}, \ref{fig:Hristo3}) 
In this example, we illustrate the effect of SO coupling on the superconducting transport. We choose $\Gamma_2^L,\Gamma_2^R,E_2 = A(1.2,1.5,-1)$, $\kappa, \gamma_L,\gamma_R = \sqrt{A}(0.2,0.6,0.2)$, $\Gamma_1^L, \Gamma_1^R = 1.6, 3.5$. 
As to spin-dependent parameters, we choose 
$\boldsymbol{\kappa} = \sqrt{A} SO [0, 0.8, 0]$, $\boldsymbol{\gamma}_L = \sqrt{A} SO [0,0.2,0]$, $\boldsymbol{\gamma}_R = \sqrt{A} SO [0,0.1,0]$ with $SO=1$ that gives $T_0 = 0.64, \Gamma_L = 1.1, \Gamma_R = 1.38, \Gamma = 2.48$ and a significant $\boldsymbol{\Gamma} = 0.45 \boldsymbol{y}$. As we see from the Figs. \ref{fig:normal_conductane_Perp_SO}, \ref{fig:Normal_Conductance_along} that give the traces of normal conductance, this set also illustrates a well-developed Fano resonance with antisymmetric features split in sufficiently high magnetic field.

We consider first $\boldsymbol{B} \perp \boldsymbol{\Gamma}$. In this case, the time-reversibility provides the symmetry $\phi \leftrightarrow - \phi$ that was present in all previous plots. Let us concentrate at the $0-\pi$ energy difference (Fig. \ref{fig:E_T_SO_perp}). The curve at zero magnetic field qualitatively follows the normal conductance. Upon increasing the magnetic field we see the multiple cusps that are already familiar from Figs. \ref{fig:Hristo0}, \ref{fig:Hristo1} and indicate the spin splitting and eventual zero crossing of ABS. The shape of the trace becomes more complex, and the minimum $E_T$ becomes smaller. However, it does not reach zero that is necessary for $0-\pi$ transition. 

The phase dependence of superconducting current at $B=2$ and various $E_1$ is presented in Fig. \ref{fig:I_SO_perp}. Most curves display pronounced discontinuities manifesting the zero crossings at corresponding phases. Except for this, the dependence is rather sinusoidal corresponding to moderate transmission. It looks like the current jumps between two sin-like curves of different amplitudes.

It is interesting to see 3 ABS in the plot presenting the phase dependence of ABS energies for $E_1 = -1.5$ and $B=2$ (Fig. \ref{fig:ABS_SO_perp}). The fourth state is either shifted over the gap edge to the continuous spectrum or is present very close to the edge so we cannot resolve it with accuracy of our numerics. The lowest state displays the familiar zero crossings corresponding to the current jumps.

When we change from perpendicular to parallel field (Fig. \ref{fig:Hristo3}) we do not see much change in normal conductance: the difference between the corresponding traces in Figs. \ref{fig:Normal_Conductance_along} and \ref{fig:normal_conductane_Perp_SO} does not exceed 10 \% . This is explained by the fact that the effect is of the second order in $\boldsymbol{\Gamma}$, $\propto \boldsymbol{\Gamma}^2/\Gamma^2$, and $|\boldsymbol{\Gamma}|/\Gamma \simeq 0.2$ is not so big. We also do not see much changes in $E_T$ traces (Fig. \ref{fig:E_T_SO_perp} versus Fig. \ref{fig:E_T_SO_along}).

The most prominent effect of SO coupling is the breaking of $\phi \leftrightarrow -\phi$ symmetry in magnetic field, the effect $\propto |\boldsymbol{\Gamma}|/\Gamma $ at $B \simeq \Gamma$. We see this in Fig. \ref{fig:I_SO_along} where the current-phase dependencies for $B=2$ are now shifted sin-like curves with jumps. The values of the shift vary from trace to trace, also in sign, and are $\simeq 0.2 - 0.3 $. In addition to the shifts of the sin-like curves, the positions of jumps are shifted non-symmetrically, these shifts are $\simeq 0 -0.5$.

Non-symmetry of the phase dependence of ABS energies is clearly seen in Fig. \ref{fig:ABS_SO_along} that is done at the same parameters as Fig. \ref{fig:ABS_SO_perp}. Also, beside shift, the energy first ABS is significantly affected by the direction of the magnetic field. A fine detail is the crossing of the second and the third ABS near $\phi \approx 1$. It may seem that in the presence of SO coupling all level crossings shall be avoided, since spin is not a good quantum number. However, since $\boldsymbol{\Gamma}$ is the only  spin vector in our model, for the particular case $\boldsymbol{B} \parallel \Gamma$ the projection of spin on $\boldsymbol{B}$ is a good quantum number and the levels of different projections may cross.

\subsection{Conclusions theory part}
\label{sec:conclusions}
To conclude, we have developed and presented a model that accurately describes normal and superconducting transport for a situation where a high transmission in a transport channel is accompanied by propagation via a resonant localized state. The motivation came from the experimental observation of a pair of $0-\pi$ transitions separated by a small interval in the gate voltage, and the model explains the main features observed at semi-quantitative level. In addition, we gave several examples not immediately related to the experiment to illustrate the rich parameter space of the model.  The accurate characterization of normal transport in the experimental setups to choose the model parameters and taking into account more resonant states should bring the agreement between experiment and theory to quantitative level.

\section{Calculation with a finite $U$ in the first dot}
\begin{figure}[!t]
\centering
\includegraphics[width=0.65\linewidth]{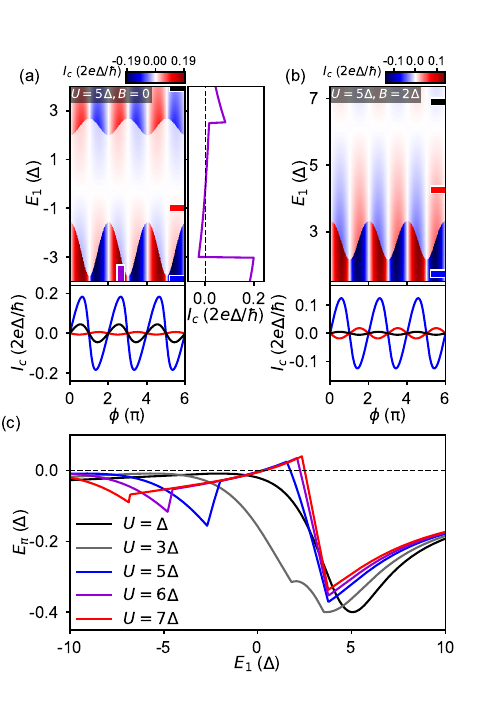}
\caption{The situation with a finite $U$ in the first dot. (a) $I_c$ as a function of $\phi$ and $E_1$ for $U=5\Delta$ and $B=0$. Three linecuts (blue, red and black marker) are shown in the bottom panel, and a linecut (purple marker) is shown in the right panel. (b) Analogous to (a), but for $B=2\Delta$. (c) Ground-state energy difference $E_\pi=E(\phi=0)-E(\phi=\pi)$ as a function of $E_1$ for $B=0$ and different charging energies $U$. $E_\pi>0$ indicates $\pi$-shifted CPR.}\label{FigWithFiniteU}
\end{figure}

Here, we consider the case with finite charging energy $U$ in the first dot. For $U=5\Delta$, $I_c$ is calculated as a function of $\phi$ and $E_1$ for Zeeman terms $B=0$ (Fig. \ref{FigWithFiniteU}(a)) and $B=2\Delta$ (Fig. \ref{FigWithFiniteU}(b)). A $\pi-$region appears already for $B=0$ (zero magnetic field), and it expands for $B=2\Delta$. $I_c$ amplitude is suppressed inside the $\pi$-region and asymmetrically modulated in the two neighboring $0-$regions, as emphasized by corresponding horizontal and vertical linecuts. Complementary to Fig. 4(e), in Fig. \ref{FigWithFiniteU}(c) the ground-state energy difference $E_\pi=E(\phi=0)-E(\phi=\pi)$ is calculated as a function of $E_1$ for various charging energies $U$ and fixed Zeeman energy $B=0$. One obtains that an interval of $E_1$ in which $E_\pi>0$ appears for sufficiently large $U$, despite the absence of the Zeeman energy. This interval corresponds to a $\pi$-region in the CPR, as $E_\pi>0$ means that the ground-state energy minimum is achieved for $\phi=\pi$ rather than $\phi=0$. Moreover, this interval broadens as $U$ increases. This demonstrates that a region with $\pi-$shifted CPR can be driven by a finite on-site interaction of the localized state even without magnetic fields, and that increasing the interaction leads to its expansion.

\section{Data selection}

Our experimental data provides the evidence of magnetic field-driven $0-\pi$ transitions and $\pi$-shifted supercurrent inside narrow intervals of the electro-chemical potential of a hybrid nanowire JJ. The presented intervals of $V_{G1}$ were identified in rough gate sweeps at high parallel fields $B_z=[600,700]\,\mathrm{mT}$ by detecting $\pi$-shifted oscillations of the SQUID switching current. These $V_{G1}$ intervals were subsequently investigated in high resolution - both at high and low $B_{z}$-fields. At $B_z=0\,\mathrm{mT}$, qualitatively different scenarios - with and without $\pi$ shifts - were observed, as illustrated in Fig. 2 and Fig. 3. Besides the $V_{G1}$ intervals in these figures, there were also few other $V_{G1}$ intervals with $\pi$-shifted supercurrent at high $B_z$-fields and evolution similar to Fig. 2 and Fig. 3. Importantly, in the rough gate sweeps at high parallel fields, we also detected many narrow $V_{G1}$ intervals in which supercurrent was sharply modulated without $\pi$ shifts. Due to the absence of striking $\pi$ shifts, these intervals were not further  examined in high resolution and at low parallel fields. Relatively low occurrence of $\pi$ shifts at high parallel fields is in agreement with the findings of our theoretical model in which $\pi$-shifted supercurrent is not generically obtained for large Zeeman energies. 

Oscillations of the SQUID switching current were observed at fields exceeding $B_z=1\,\mathrm{T}$. However, the reliability of detecting switches in $V-I_b$ traces by the setup for fast switching current measurements was considerably lower for $B_z>750\,\mathrm{mT}$ - due to less sharp switches in $V$. Therefore, $I_{sw}$ could be reliably and efficiently measured up to $B_z \sim 720\,\mathrm{mT}$.

\section{Data and code availability}
The data, the code and the figures of theoretical calculations can be found at \url{https://doi.org/10.5281/zenodo.5879475}. The raw data and analyzing code of experimental measurements can be found at \url{https://doi.org/10.5281/zenodo.8364886}.